\documentclass[11pt,a4paper]{article} 
\usepackage{jcappub}
\usepackage{bm}
\usepackage{times} 
\usepackage[utf8]{inputenc}
\usepackage[english]{babel}
\usepackage[T1]{fontenc}
\usepackage{graphicx}
\usepackage{amsmath}
\usepackage{subfig}
\usepackage{hyperref}
\usepackage{color}
\usepackage{amssymb}
\usepackage[font=scriptsize,labelfont=bf]{caption}
\usepackage{appendix}
\usepackage{slashed}
\usepackage{slantsc}
\usepackage{braket}
\usepackage{mathrsfs}
\usepackage{subfig}
\usepackage{placeins}
\usepackage{fullpage}
\usepackage{sidecap}
\usepackage{float}
\usepackage{wrapfig}
\usepackage[dvipsnames]{xcolor}
\usepackage{dsfont}
\usepackage{framed}
\usepackage{lipsum}

\newcommand{\nnn}{\nonumber \\}
\newcommand{\beeq}{\begin{equation}}
\newcommand{\eneq}{\end{equation}}
\newcommand{\rbr}[1]{\left(#1\right)}
\newcommand{\sbr}[1]{\left[#1\right]}
\newcommand{\angbr}[1]{\left\langle#1\right\rangle}
\newcommand{\obs}{{\text{obs}}}
\newcommand{\texth}{{\text{th}}}
\newcommand{\rz}{{\bar r_z}}
\newcommand{\dr}{\mathrm d\bar r\,}
\newcommand\blfootnote[1]{%
  \begingroup
  \renewcommand\thefootnote{}\footnote{#1}%
  \addtocounter{footnote}{-1}%
  \endgroup
}

\begin{document}

\begin{titlepage} 

\begin{center}

\vskip 1.0 cm

{\LARGE \bf Galaxy Power Spectrum in General Relativity}

\vskip 1.0 cm

{\large Nastassia Grimm$^{a,\dagger}$,
Fulvio Scaccabarozzi$^{a}$, Jaiyul Yoo$^{a,b}$, Sang Gyu Biern$^{c}$ \\ and Jinn-Ouk Gong$^{d}$
}

\vskip 0.5cm

{\it
$^{a}$Center for Theoretical Astrophysics and Cosmology,
Institute for Computational Science, University of Z\"urich, Winterthurerstrasse 190, CH-8057, Z\"urich, Switzerland
\vspace{0.2 cm}
\\
$^{b}$Physics Institute, University of Z\"urich, Winterthurerstrasse 190, CH-8057, Z\"urich, Switzerland
\vspace{0.2 cm}
\\
$^{c}$Optotune, Bernstrasse 388, CH-8953, Dietikon, Switzerland
\vspace{0.2 cm}
\\
$^{d}$Korea Astronomy and Space Science Institute, Daejeon 34055, Korea
}
\blfootnote{$^\dagger$E-mail: ngrimm@physik.uzh.ch}

\vspace{1 cm}

\today

\vskip 1.0 cm

\end{center}

\vspace{1.5cm}
\hrule \vspace{0.3cm}
\noindent {\sffamily \bfseries Abstract} \\[0.1cm]
We present the galaxy power spectrum in general relativity. 
Using a novel approach, we derive the galaxy power spectrum taking into account all the relativistic effects in observations. 
In particular, we show independently of survey geometry that relativistic effects yield \textit{no} divergent terms (proportional to $k^{-4}P_m(k)$ or $k^{-2}P_m(k)$ on all scales) that would mimic the signal of primordial non-Gaussianity. This cancellation of such divergent terms is indeed expected from the equivalence principle, meaning that any perturbation acting as a uniform gravity on the scale of the experiment cannot be measured. We find that the unphysical infrared divergence obtained in previous calculations occurred only due to not considering all general relativistic contributions consistently. 
Despite the absence of divergent terms, general relativistic effects represented by non-divergent terms alter the galaxy power spectrum at large scales (smaller than the horizon scale). In our numerical computation of the full 
galaxy power spectrum, we show the deviations from the standard redshift-space power spectrum due to these non-divergent corrections. We conclude that, as relativistic effects significantly alter the galaxy power spectrum at $k\lesssim k_{eq}$, they need to be taken into account in the analysis of large-scale data.
\vskip 10pt
\hrule
\vspace{0.6cm}
\end{titlepage}

\pagebreak

\noindent\hrulefill
{\linespread{0.75}\tableofcontents}
\noindent\hrulefill

\section{Introduction}

The next generation of galaxy surveys \cite{Laureijs:2011gra,Amendola:2016saw,Abate:2012za,Abell:2009aa,Aghamousa:2016zmz,Carilli:2004nx} is going to explore the large-scale structure of the universe with unprecedented precision in their measurements and parameter estimation.
By measuring the positions of millions of galaxies, these surveys will probe the distribution of galaxies at high redshifts on very large scales, where the relativistic effects are more important.
In order to extract physical information from this map, different observables are computed, such as the galaxy correlation function and power spectrum. However, the theoretical expression for the galaxy number density used for data analysis usually takes into account only matter density fluctuations and redshift-space distortions, while ignoring relativistic effects (for standard literature on galaxy clustering and redshift-space distortions using Newtonian dynamics, see e.g.~\cite{kaiser, RSD1,RSD2,RSD3,RSD4,RSD5}). This standard expression provides a reasonably good approximation to what we observe, which is, however, not sufficiently accurate to interpret high-precision measurements from upcoming surveys.

The impact of relativistic effects on the galaxy correlation function and power spectrum has already been studied in previous works. 
The theoretical prediction for the galaxy correlation function, including all relativistic effects, has been recently presented in \cite{Fulvio,Vitto,Tansella:2018sld}. Given the expression for the observed galaxy number density, it can be derived directly in terms of observable quantities, simply connecting two points on the sky.
The galaxy power spectrum, however, is mathematically more involved due to difficulties arising from the non-local nature of the Fourier transform and the non-flat sky (see~\cite{YooDesjacques} for using the spherical power spectrum on the sky). 

Taking into account the kinematic Doppler effect in redshift-space, Kaiser~\cite{kaiser} pioneered the study on the relation between the physical and the observed quantities in Newtonian dynamics and presented the anisotropic galaxy power spectrum in the observed coordinates, i.e.~the redshift space. Extending this concept to a fully general relativistic study, a gauge-invariant expression for the galaxy number density was first obtained in~\cite{Yoo:2009,Yoo:2010}. This expression contains general relativistic contributions evaluated at the source position, as well as at the observer position and along the line-of-sight such as the gravitational lensing contribution. Based on this solution, the corresponding galaxy power spectrum was derived, and its detection significance was quantified in \cite{Yoo:2010,Yoo:2012,Jeong}. However, because of the difficulties associated with the Fourier decomposition, these works included only the relativistic effects at the source position in their analysis of the power spectrum, while ignoring those at the observer position and along the line-of-sight. It was shown in~\cite{Fulvio} that
omitting these contributions leads to a gauge-dependent expression for the galaxy number density which does not correctly describes observations.  As a consequence, the galaxy power spectrum obtained in \cite{Yoo:2010,Yoo:2012,Jeong} using this incomplete expression diverges on super-horizon scales. In literature, this result of infrared divergence was taken seriously in the interpretation of future large-scale power spectrum measurements, as the purely relativistic effects appear to generate signals on large scales similar to those from the local-type primordial non-Gaussianity~\cite{Baldauf2011, Bartolo2011, Xia2011, Hamaus2011, Jeong2012, Sawangwit2012, Giannantonio2012}.
In this work we show that, when all the contributions neglected in previous works are taken into account, all infrared-divergent terms in the power spectrum in fact cancel out. Indeed, this cancellation of the infrared divergence is expected from the compatibility of the expression for the observable with the equivalence principle (see \cite{Fulvio,Jeong,SG divergence,cosmic rulers}). Therefore, at linear order in perturbations there are \textit{no} divergent terms arising from general relativistic effects that would mimic the signal of primordial non-Gaussian signature . 

In this work, we first investigate the relation between the theory and observed galaxy power spectra, which provides the possibility to correctly analyze observed data. Then, we focus on the derivation of the theory power spectrum in general relativity. We present a simple method to compute the galaxy power spectrum from the variance of the observed galaxy fluctuation, which allows to take into account all relativistic effects, including those evaluated at the observer position and along the line-of-sight direction. With this method, we derive the expression for the full anisotropic galaxy power spectrum. As mentioned above, we show that when all terms are taken into account, the theory power spectrum is devoid of the infrared divergence claimed in previous works \cite{Yoo:2010,Yoo:2012,Jeong}. 
Furthermore, we numerically investigate the behavior of the theory power spectrum on large scales, where the relativistic effects manifest themselves and cause deviations from the standard redshift-space prediction given by the Kaiser formula. 

While a correct description of the relativistic effects is essential to understand the clustering of galaxies on large scales (from the theoretical point of view), in order to obtain the detection significance one has to compute the observed power spectrum. Obviously, a very large survey volume is needed to observe the impact of relativistic effects on the galaxy power spectrum. Here, we do two important steps towards that direction: We clarify what is actually observed and provide the correct interpretation of the galaxy power spectrum commonly computed in theory, and we derive the correct theoretical expression considering all relativistic effects.

The organization of this paper is as follows. First, we introduce the expression for the observed galaxy fluctuation with all required ingredients in sections~\ref{metric and solutions} and \ref{Observed Galaxy Number Density}. Then, we discuss the distinction between the observed galaxy power spectrum and the theory one in section~\ref{th vs obs}, providing the mathematical relation between the two quantities. We then study the theory power spectrum in section~\ref{PS}, where we derive its analytical expression (section~\ref{contribution to PS}) and discuss how the issue concerning the infrared divergence is resolved (section~\ref{IRdiv}). Furthermore, in section~\ref{RSDeL}, we consider the galaxy power spectrum when accounting for the standard redshift-space distortion and standard lensing only. In section~\ref{numerical}, we compute the theory power spectrum numerically and discuss the impact of general relativistic effects on the monopole (section~\ref{monopole}), quadrupole (section~\ref{quadrupole}) and hexadecapole (section~\ref{hexadecapole}) of the power spectrum. We conclude with a summary and discussion in section~\ref{conclusion}. In appendix \ref{A}, we provide the expressions for the cross power spectra of the different contributions to the galaxy fluctuation.

\section{Preliminaries}

In this section, we first introduce in section~\ref{metric and solutions} the perturbation variables needed for the expression of the observed galaxy number density, with the solutions for their time evolution in $\Lambda$CDM. Next we present in section~\ref{Observed Galaxy Number Density} the gauge-invariant expression for the observed galaxy number density fluctuation, which will be referred to as the observed galaxy fluctuation. Finally, in section~\ref{th vs obs}, we define the \textit{theory} Fourier modes of the observed galaxy fluctuation and discuss the subtleties associated with them.
 
\subsection{Metric convention and $\Lambda$CDM solutions for scalar perturbations}\label{metric and solutions}

We adopt a spatially flat Friedmann-Robertson-Walker (FRW) metric for our theoretical description of the background universe. 
For an inhomogeneous universe, we consider only linear-order scalar perturbations and a pressureless medium (dark matter and baryons on large scales).
We choose the conformal Newtonian gauge:
\begin{equation}
\mathrm ds^2=-a^2(1+2\Psi)\mathrm d\eta^2 + a^2(1-2\Psi)\mathrm dx^2\,,
\end{equation}
where $\eta$ is the conformal time, $x^i$ are the Cartesian coordinates, $a(\eta)$ is the scale factor and $\Psi(\eta,\boldsymbol{x})$ is the linear-order gravitational potential. 
In this space-time, the observer moves with time-like four-velocity $u^{\mu}\equiv a^{-1} (1 - \Psi , V^i)$, where the spatial component can be expressed in terms of a scalar perturbation $v(\eta,\boldsymbol{x})$ as $V_i \equiv - \partial_i v$. 

The observer identifies the position of a source galaxy by measuring the redshift $z$ and the angular direction $\boldsymbol{\hat n}$ of the incoming photons. 
Given these quantities, the observed source position $x^\mu=(\bar\eta_z, \bar r_z  \hat n^i)$ can be computed by using the distance-redshift relation in a homogeneous universe,
\begin{equation}
\bar r_z=\bar \eta_o - \bar \eta_z = \int_0^{z} \frac{\mathrm dz'}{H(z')} \,,
\end{equation}
where $H$ is the Hubble parameter and a bar indicates that the coordinates are computed in the background universe at the observer ($o$) and the source (at redshift $z$) positions.
However, the real position of the source galaxy is different from the one inferred in the background universe, because the photon propagation is affected by the inhomogeneities.

Before introducing the theoretical prediction for the galaxy number density, we need the $\Lambda$CDM solutions for the scalar perturbations that enter the expression.
In a universe with a pressureless medium, all Fourier modes at linear order grow at the same rate and the time dependence of the scalar perturbations in the conformal Newtonian gauge can be expressed in terms of the growth function $D$ of the matter density contrast,
\beeq
\delta_m(\eta,\boldsymbol{x})= [ D(\eta)/D(\bar\eta_o) ] \delta_m(\boldsymbol{x})\,, \label{matterpert}
\eneq
where $\delta_m(\boldsymbol{x})$ is the spatial configuration at the present time.
Using the conservation of energy and momentum in a $\Lambda$CDM universe, one can derive the evolution equation for the linear growth function $D(\eta)$ in terms of the scale factor
\begin{equation}
	\frac{\mathrm d^2 D}{\mathrm da^2} + (2-\Omega_m)\frac{3}{2a}\frac{\mathrm d D}{\mathrm da}-\frac{3}{2a^2} D = 0 \,,
\end{equation}
and the solution is
\begin{equation}\label{D1}
	D(a) \propto a \, _2 F_1\bigg[\frac{1}{3},1,\frac{11}{6},-\frac{a^3}{\Omega_m}(1-\Omega_m)\bigg]\,,
\end{equation}
where $_2 F_1$ is the hypergeometric function and $\Omega_m$ is the matter density today.
Furthermore, all the other perturbation variables can be expressed in terms of the initial curvature perturbation $\zeta(\boldsymbol{x})$ in the comoving gauge as 
\begin{equation}\label{rel}
\begin{split}
&\Psi(\eta,\bm x) = D_\Psi(\eta) \zeta(\bm x) \,, \qquad v(\eta,\bm x) = - D_V(\eta) \zeta(\bm x) \,, \qquad V_{i}(\eta,\bm x) = D_V(\eta) \partial_{i}\zeta(\bm x)  \,,
\end{split}
\end{equation}
where the growth functions $D$ and $D_\Psi$ are dimensionless, while $D_V$ has a dimension of length. As described in~\cite{Yoo:2016tcz}, the time-dependent growth functions $D_\Psi$ and $D_V$ are related to $D$ through 
\begin{equation}\label{rel 2}
D_\Psi = \frac{1-\Sigma}{\Sigma} \,,  \qquad  D_V = \frac{1}{a H \Sigma}\,,  \qquad  \Sigma \equiv 1 + \frac{3}{2}\frac{\Omega_m}{f} \,, \qquad f \equiv \frac{\mathrm d \ln D}{\mathrm d \ln a} \,,
\end{equation}
and, furthermore, $D_\Psi$ and $D_V$ satisfy the following equations:
\begin{equation}\label{relgf}
 D_\Psi = \mathcal H D_V -1 = - \mathcal H D_V - D_V' = -\frac{1}{2}(D_V'+1)\,, \qquad \int_0^{\bar r_z}\mathrm d\bar r\, D_\Psi = \frac{1}{2}(D_V - {D_{V}}_o - \bar r_z)  \,.
\end{equation}
Here, $\mathcal H=aH$ is the conformal Hubble parameter and a prime denotes the derivative with respect to conformal time.
Finally, the relation between the spatial configuration $\delta(\boldsymbol{x})$ of the density contrast and the curvature perturbation $\zeta(\boldsymbol{x})$ today is given by 
\begin{equation}\label{zeta}
\zeta(\bm x) = C \Delta^{-1} \delta_m(\bm x) \,.
\end{equation}
Since $\zeta$ is time-independent in the late universe, $C \equiv -  \mathcal H^2  \Sigma f D$ is a constant and has dimension of $r^{-2}$. This completely determines the relation of all the perturbation variables to the density contrast at $z=0$. From now on we use the notation $D$ for the growth factor normalized at the present epoch.

Finally, we point out the solutions to the growth functions and related quantities in an Einstein-de-Sitter universe, as this will be useful in subsequent sections to analyze high-redshift behavior. In such a universe, the matter fluctuation grows with the scale factor:
\begin{equation}
D(a)=a~,\qquad\qquad f=1~,\qquad\qquad \Sigma={5\over2}~,\qquad\qquad
D_\Psi=\frac35~.
\end{equation}
The Friedmann equations in the Einstein-de~Sitter universe yield
\begin{equation}
a(\eta)=\left({\mathcal H_o\eta\over2}\right)^2~,\qquad
\mathcal H=H_o a^{-1/2}~,\qquad
 \mathcal H'=-{\mathcal H^2\over2}=-{H_o^2\over2a}~,\qquad
D_V={2a^{1/2}\over5H_o}~.
\end{equation}

\subsection{Observed galaxy number density}\label{Observed Galaxy Number Density}

The observed galaxy number density $n_g$ is obtained by counting the number of galaxies within the observed volume, which is the volume in a homogeneous universe within the observed redshift interval and the observed solid angle. However, in the presence of inhomogeneities in the universe, the observed volume does not correspond to the physical one occupied by the observed galaxies. Such difference contributes to the fluctuation $\delta_g$ in the galaxy number density that can be described as $n_g \equiv \bar n_g (\bar \eta_z)(1+\delta_g)$, where $\bar n_g$ is the mean density. Since the galaxy number density is a physical observable, the theoretical expression of $\delta_g$, derived as a function of the observed redshift and angles, is gauge-invariant \cite{Yoo:2009,Yoo:2010,Bonvin:2011bg,Challinor:2011bk,Jeong}. 

\begin{figure}
	\centering
		\includegraphics[width=\textwidth]{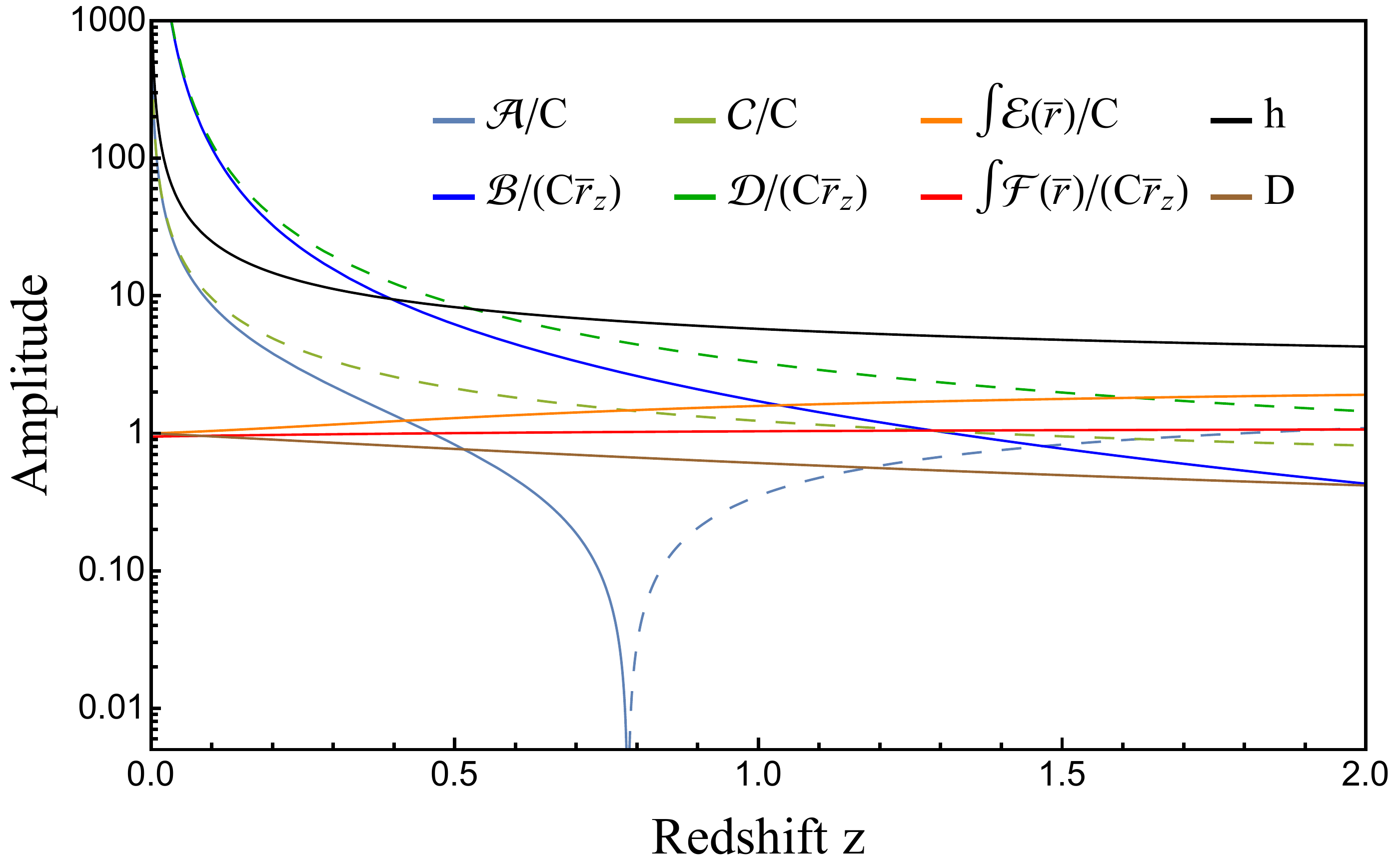}
		\caption{Redshift dependence of the functions $\mathcal A$, $\mathcal B$, $\mathcal C$, $\mathcal D$, $\mathcal E(\bar r)$, $\mathcal F(\bar r)$, $h$ and $D$ defined in equations~\eqref{ds}--\eqref{calI}, \eqref{h} and~\eqref{matterpert}, where dashed lines represent negative values. In order to plot dimensionless quantities we divided all functions, except for $h$ and $D$, by the constant $C$ (defined in equations~\eqref{zeta}) and $\mathcal B$, $\mathcal D$, $\mathcal F(\bar r)$ additionally by the comoving distance $\bar r_z$. Note that the integrals of $\mathcal E(\bar r)$ and $\mathcal F(\bar r)$ are performed along the line-of-sight comoving distance from zero to $\bar r_z$ evaluated at the redshift value along the horizontal axis.}
	\label{z-functions}
\end{figure}

We split the expression of the observed galaxy fluctuation into the local contributions evaluated at the source ($s$) and the observer  ($o$) positions, and the non-local ($nl$) contributions: $\delta_g(z,\boldsymbol{\hat n}) = \delta_s(z,\boldsymbol{\hat n}) + \delta_o(z,\boldsymbol{\hat n}) + \delta_{nl}(z,\boldsymbol{\hat n})$, where we use the term ``non-local'' to refer to the contributions arising from the line-of-sight integration as opposed to those localized at the observer or the source position.
Written as a function of the observed redshift and angles, the expression in the conformal Newtonian gauge is given by
\begin{align} 
\label{dg}
	\delta_s &= b  \delta_m - e \mathcal H v + \big[ 4 -h(z) \big] (  V_{\parallel} - \Psi )  -\frac{1}{\mathcal H} ( \partial_{\parallel}V_{\parallel}  - V_{\parallel}' - \partial_{\parallel}\Psi - \Psi') \,,
\\ 
	\delta_o &= - \bigg[ 3 - h(z) +  \frac{2}{\bar r_z \mathcal H_o } \bigg] \mathcal H_o v_o + \big[ 3 - h(z) \big]  \Psi_o + \big[ h(z) -1 \big] {V_{\parallel }}_o  \,,
\\ \label{nl}
	\delta_{nl} &= 
	 \int_0^{\bar r_z}\mathrm d\bar r\, \bigg[ \frac{4}{\bar r_z} \Psi
	+ 2 \big[ h(z) - 3 \big]  
	    {\Psi}'  -2  \bigg(\frac{\bar r_z - \bar r}{\bar r_z \bar r}\bigg)\hat\nabla^2 \Psi  \bigg]  \,,
\end{align}
where $b$ is the galaxy bias, $\delta_m$ is the matter density contrast in the comoving gauge, $e \equiv\mathrm  d \ln \bar n_g / \mathrm d \ln (1+z)$ is the evolution bias ($e=3$ for a constant comoving number density), $V_{\parallel}\equiv\hat n^i V_i$ is the line-of-sight velocity, $\partial_{\parallel}\equiv\hat n^i \partial_i$ is the derivative along the line of sight, $\hat\nabla^2$ is the angular Laplacian, related to the 3D Laplacian $\hat\nabla^2=\bar r^2 \Delta - 2\bar r\, \partial_{\bar r} - \bar r^2 \partial_{\bar r}^2\,$, and $\bar r$ is the radial coordinate corresponding to the comoving distance. Furthermore, we defined the redshift-dependent function 
\beeq
h(z) \equiv e + \mathcal H'/\mathcal H^2  + 2 / (\bar r_z \mathcal{H})\,, \label{h}
\eneq 
for ease of notation.\footnote{Note that the definition of $h(z)$ here corresponds to $3-h(z)$ with $h(z)$ defined as in \cite{Fulvio}, where the evolution bias $e$ is instead called $e_z$.} As visible in figure~\ref{z-functions}, this function diverges at low redshift due to the factor $1/\bar r_z$ in the third term. At higher redshift, this term contributes less, as it goes to zero in a matter-dominated universe, while the second term converges to $-0.5$, implying that the total $h$ converges to $e-0.5$ .

The fluctuation $\delta_s$ at the source position includes the standard redshift-space distortion contribution,
\begin{equation} 
 \delta_z(z,\boldsymbol{\hat n})\equiv b  \delta_m -\frac{1}{\mathcal H}  \partial_{\parallel}V_{\parallel} \,, \label{deltaz}
\end{equation} 
and the non-local fluctuation $\delta_{nl}$ also includes the standard gravitational lensing effect,
\begin{equation}
\delta_L(z,\boldsymbol{\hat n})\equiv  -2  \int_0^{\bar r_z} \mathrm d\bar r\,  \bigg(\frac{\bar r_z - \bar r}{\bar r_z \bar r}\bigg)\hat\nabla^2 \Psi \,. \label{deltaL}
\end{equation} 
Using equations~\eqref{rel}$-$\eqref{zeta}, and the relation $\hat\nabla^2 e^{i k \bar r\,\mu_k} =- k \bar r\, \big[2i \mu_k + k \bar r (1-\mu_k^2)\big] e^{i k \bar r \mu_k}$, we can re-arrange the three contributions $\delta_s$, $\delta_o$, $\delta_{nl}$ to the observed galaxy fluctuation as
\begin{align}\label{ds}
		 \delta_s (z,\boldsymbol{\hat n})  &= \int \frac{\mathrm d^3 k}{(2\pi)^3} e^{i k \bar r_z\mu_k} 
		  \sbr{   \mathcal A \,  \frac{1}{k^2} 
		 + 
		\mathcal B  \, \frac{i \mu_k}{k} +    D \big(b
		+  f  \mu_k^2 \big) } \delta(\bm k)  \,,
\\
		 \delta_o (z,\boldsymbol{\hat n})  &= \int \frac{\mathrm d^3 k}{(2\pi)^3}\sbr{  
		 \mathcal C  \frac{1}{k^2} 
		   + \mathcal D \frac{i \mu_k}{k}} \delta(\bm k)  \,,
\\\label{dnl}
		 \delta_{nl} (z,\boldsymbol{\hat n})  &= \int \frac{\mathrm d^3 k}{(2\pi)^3} \int_0^{\bar r_z}\mathrm d\bar r\, e^{i k \bar r \mu_k} \sbr{ 
		 \mathcal E(\bar r)  \frac{1}{k^2} 
+ \mathcal F(\bar r) \,  \frac{i \mu_k}{k}
 +  \mathcal G(\bar r) \,   (1 - \mu_k^2) }  \delta(\bm k)  \,,
\end{align}
where $\delta(\bm k)$ is the Fourier transform of the matter density contrast today, $\mu_k \equiv\boldsymbol{\hat k}\cdot \boldsymbol{\hat n} $ is the cosine angle between the wavevector and the line-of-sight direction, and we defined the following functions of time (or redshift):
\begin{align} \label{calA}
&	\mathcal A  \equiv \bigg(- 2\,[h-3] \mathcal HD_V + \frac{2}{\bar r_z } D_V - 5 + h \bigg) C  \,,
\qquad
	\mathcal B  \equiv   \big[ h-3 \big] D_V   C  \,,
	\\
&	\mathcal C  \equiv  \bigg(  [3-h](1-2\mathcal H_o D_{Vo}) - \frac{2}{\bar r_z}D_{Vo} \bigg)C \,,
\qquad
	\mathcal D  \equiv - \big[ h -1 \big]  D_{Vo} C \,,
	\\\label{calI}
&	\mathcal E(\bar r)  \equiv \bigg( - \frac{4}{\bar r_z}  D_\Psi(\bar r)  
		- 2 \big[ h - 3 \big]    D_\Psi'(\bar r)  \bigg) C  \,,
\qquad 
		\mathcal F(\bar r)  \equiv    -  4  \bigg(\frac{\bar r_z - \bar r}{\bar r_z }\bigg)    D_\Psi(\bar r)  C \equiv \frac{2}{\bar r} \,\mathcal G (\bar r)  \,.
\end{align}

These functions are all proportional to the constant $C$, i.e.~the initial fluctuation amplitude. Their dependencies on redshift (see figure~\ref{z-functions}) are determined by their relations to $h$, $\mathcal H$, $\bar r_z$, $D_V$, $D$ and $f$. First, note that the functions $\int \mathcal E(\bar r)/C$ and $\int\mathcal F(\bar r)/(C\bar r_z)$ are both almost constant. Indeed, if we assume that $D_\Psi$ is constant and thus $D'_\Psi=0$, as in the case of a matter-dominated universe, the integrals are equal to $-4D_\Psi$ and $-2D_\Psi$, respectively. The fact that these functions are, however, not exactly constant is due to the changing potential at low redshift, where the cosmological constant has become important. The functions $\mathcal A/C$, $\mathcal B/C$, $\mathcal C/C$ and $\mathcal D/(C\bar r_z)$, however, all diverge at $z\rightarrow 0$, which is an immediate consequence of the factors $1/\bar r_z$ appearing in their definitions. To analyze the functions' behavior at high redshifts, where the cosmological constant has little significance, we can use the Einstein-de Sitter solutions presented in section~\ref{metric and solutions} and the fact that $h\rightarrow 2.5$ for $e=3$. The matter growth function $D$ is normalized to $D=1$ at $z=0$ and follows the behavior $D\propto a$ at high redshifts, as already discussed in section~\ref{metric and solutions}. For $\mathcal A/C$, the contribution of the term $\sim 1/\bar r_z$ converges to zero at high redshifts, while the sum of the other terms converges to $-2.1$. The function $\mathcal B/C$ is proportional to $a^{1/2}$ in a matter-dominated universe, and thus approaches zero. Similarly, the function $\mathcal C/C$ converges to $3-h\approx 0.5$, as the other terms contain a factor $a^{1/2}$. Finally, the function $\mathcal D/C$ converges to $-1.5 {D_V}_o/\eta_o\approx -0.24$.

Furthermore, note that if we set $\mathcal A=\mathcal B =0$, then $\delta_s $ reduces to the standard redshift-space distortion galaxy fluctuation $\delta_z$, and, if $\mathcal E=0$, then $\delta_{nl}$ reduces to the standard lensing contribution to the galaxy fluctuation $\delta_L$. As we shall see in section~\ref{contribution to PS}, the expressions in equations~\eqref{ds}$-$\eqref{dnl} are more practical to compute the power spectra of different contributions to the observed galaxy fluctuation. We will derive the galaxy power spectrum as the sum of such contributions and also show that there is no divergence on large scales.

\begin{figure}[h]
	\centering
		\includegraphics[width=0.8\textwidth]{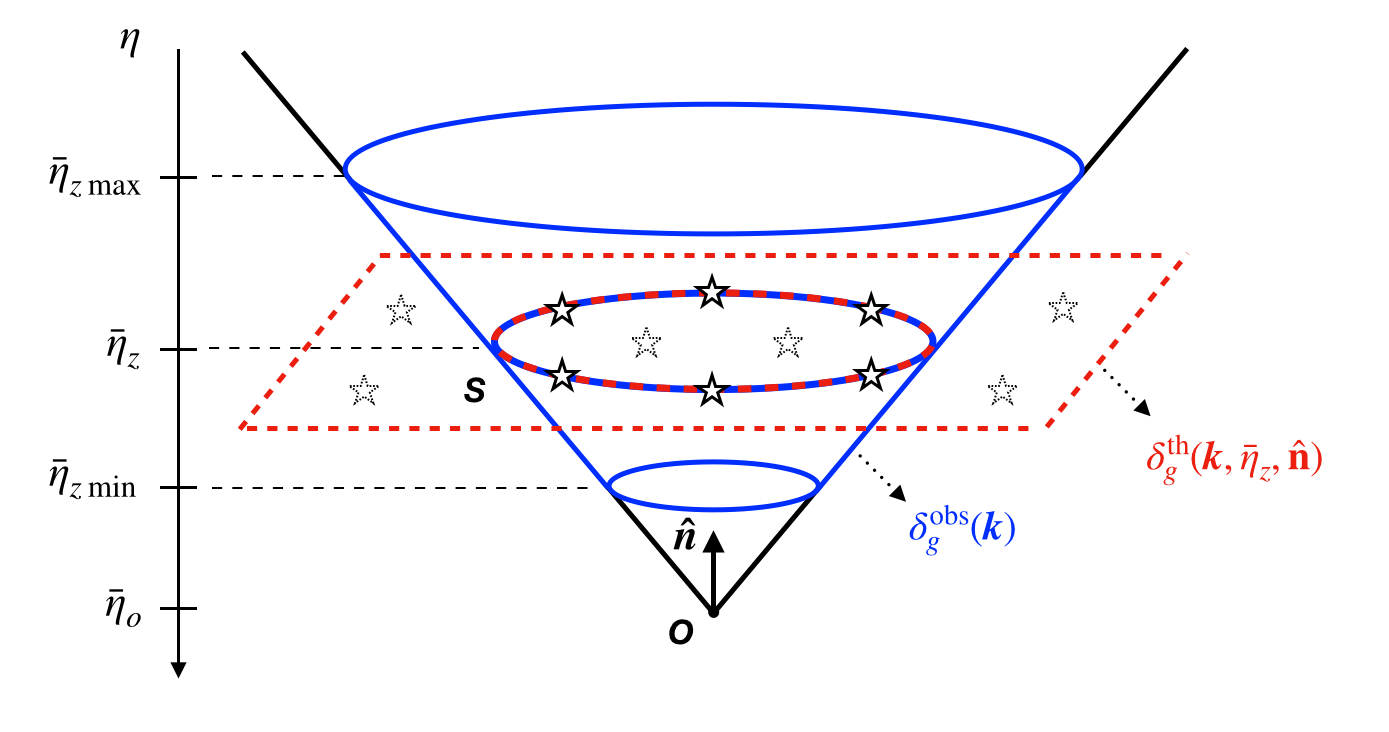}
		\caption{Sketch of the observational light-cone volume and the hypersurface of simultaneity; The observer is identified at the background time coordinate $\bar \eta_o$, and we consider the source field at one (observed) redshift slice~$z$ (time coordinate $\bar \eta_z$). The (survey) light-cone volume drawn in blue corresponds to the observed hypersurface on which the observed Fourier mode $\delta_g^{\text{obs}}(\boldsymbol{k})$ of $\delta_g(\boldsymbol{x}_\text{obs})$ is defined. The (spatially infinite) hypersurface of simultaneity at the source redshift drawn in red, in which the theory Fourier modes $\delta_g^\text{th} (\boldsymbol{k};\bar\eta_z,\boldsymbol{\hat n})$ of $\delta_g(\boldsymbol{x}_\text{obs})$ are defined. Note that the intersection (blue and red) with the light-cone is the only part of the hypersurface of simultaneity to which we have access in observation. We assume that the survey ranges over redshift $z\in[z_{\rm min},z_{\rm max}]$.}
	\label{light-cone}
\end{figure}

\subsection{Theory Fourier mode and theory power spectrum}
\label{th vs obs}

In this section, we discuss the relation between the \textit{theory} power spectrum that we compute in section~\ref{PS} and the \textit{observed} power spectrum measured in galaxy surveys. As we shall see, the key difference between the two quantities is the hypersurface used to determine the respective Fourier modes: The theory power spectrum is defined on a (infinite) hypersurface of simultaneity shown as red dotted in figure~\ref{light-cone}, while the observed hypersurface is given by the past light-cone volume within the survey boundary shown as blue solid. This light-cone volume, which the observers use to determine the observed power spectrum, encompasses many time slices, while the theory Fourier modes and power spectrum are defined in a hypersurface of simultaneity determined by a constant time coordinate.

In the survey light-cone volume~$V$, the observed Fourier mode~$\delta_g^{\text{obs}}(\boldsymbol{k})$ is simply obtained by taking a Fourier transformation of the observed galaxy fluctuation~$\delta_g(\boldsymbol{x}_{\rm obs})$ on the finite volume $V$, as
\begin{equation}
\delta_g^{\text{obs}}(\boldsymbol{k})\equiv\int_V\mathrm d^3x_{\rm obs}\,e^{-i\boldsymbol{k}\cdot
\boldsymbol{x}_{\rm obs}}\delta_g(\boldsymbol{x}_{\rm obs})\,.
\end{equation}
In contrast, the theory Fourier mode~$\delta_g^{\rm th}(\boldsymbol{k};\eta_z,\boldsymbol{\hat n})$ is, regardless of observational accessibility, defined on a hypersurface of simultaneity set by the observed redshift~$z$ (or any time coordinate) as
\begin{equation}
\label{eq:theory}
\delta_g(\boldsymbol{x}_{\rm obs})=\int{\mathrm d^3k\over(2\pi)^3}\,e^{i\boldsymbol{k}\cdot\boldsymbol{x}_{\rm obs}}\delta_g^{\rm th}
(\boldsymbol{k};\bar\eta_z,\boldsymbol{\hat n})\,, 
\end{equation}
where the observed position on the light cone is in a FRW coordinate,
\begin{equation}
\boldsymbol{x}_{\mathrm{obs}}\equiv \bar r_z{\boldsymbol{\hat n}}\,.
\end{equation}
In this work, our focus is on the theory Fourier mode (or the theory power spectrum). However, it is in fact not a direct observable, but only related to the observed power spectrum as we explain below. In literature, the distinction between the theory and observed power spectra is rarely made. Indeed, in theoretical work, we are mainly used to theory Fourier modes and power spectra. Quantities such as the matter density $\delta_m(\boldsymbol{k};\eta)$, the peculiar velocity $V_i(\boldsymbol{k};\eta)$ and the gravitational potential $\Psi(\boldsymbol{k};\eta)$ are independently of observations defined on a hypersurface of simultaneity specified by the time coordinate $\eta$, and evolved according to the solutions of the Einstein equation in section~\ref{metric and solutions}. To compute the matter power spectrum today, the Boltzmann code is often applied, evolving the matter power spectrum $P_m(k;\eta)$ in time and hence corresponding to a non-observable theory power spectrum defined within a constant-time hypersurface. 

Despite not being directly observable, the theory power spectrum is, as we discuss further in section~\ref{PS}, a highly suitable tool for theoretical investigations of the observed galaxy fluctuation $\delta_g(\boldsymbol{x}_{\rm obs})$ in Fourier space. As one advantage, it is independent of the specifications of the survey geometry, which complicates the computation of the observed power spectrum in real surveys.
Here, for completeness, we describe how the observed power spectrum can be derived from the theory one when a specific survey geometry is given. First, note that the observed Fourier mode can be related to the theory Fourier mode as
\begin{align}
\delta_g^{\rm obs} (\boldsymbol{k}) &\equiv \int_{V}\mathrm d^3 x_\text{obs} \, e^{- i\boldsymbol{k}\cdot\boldsymbol{x}_\text{obs}}\delta_g (\boldsymbol{x}_\text{obs}) = \int_{V}\mathrm d^3 x_\text{obs} \, e^{- i\boldsymbol{k}\cdot\boldsymbol{x}_\text{obs}} \int  \frac{\mathrm d^3k'}{ (2\pi )^3}  e^{i \boldsymbol{k}'\cdot\boldsymbol{x}_\text{obs}} \delta^\text{th}_g( \boldsymbol{k}' ; \bar \eta_z ,\boldsymbol{\hat n}) \nnn
&= \int  \frac{\mathrm d^3k'}{ (2\pi )^3} \int_{V}\mathrm d^3 x_\text{obs} \,e^{ i(\boldsymbol{k}'-\boldsymbol{k})\cdot\boldsymbol{x}_\text{obs}} \delta^{\text{th}}_g( \boldsymbol{k}' ; \bar \eta_z ,\boldsymbol{\hat n})  \,.
\end{align} 
Due to the dependence on the angular and time coordinates, the theory Fourier mode
cannot be pulled out of the integration over the light-cone volume~$V$.
Only if the survey volume is shallow in redshift depth and narrow in angle, it can be pulled out
of the integration, and the volume integration can be approximated as a
Dirac delta function to yield $\delta_g^{\rm obs}(\boldsymbol{k})\approx\delta_g^{\text{th}}
(\boldsymbol{k};\bar\eta_z,\boldsymbol{\hat n})$. In general, the observed Fourier mode
is a convolution of the theory Fourier mode over multiple hypersurfaces
set by the redshift range of the survey and also the survey geometry encoded
in the integration range~$V$. The observed power spectrum is then related to
the theory power spectrum as
\begin{align}
 &(2\pi)^3 \delta^D (\boldsymbol{k}_1-\boldsymbol{k}_2)P_{\obs}(\boldsymbol{k_1})\equiv\angbr{ \delta_g^\obs ( \boldsymbol{k}_1) {{\delta_g^{\obs\,\ast}}}( \boldsymbol{k}_2)} \nnn
 &= \int \frac{\mathrm d^3k}{\left(2\pi\right)^3} \int \frac{\mathrm d^3k'}{\left(2\pi\right)^3} \int_{V}\mathrm d^3 x_1 \int_{V} \mathrm d^3 x_2 \, e^{i (\boldsymbol{k}-\boldsymbol{k}_1)\cdot\boldsymbol{x}_1} e^{-i(\boldsymbol{k}'-\boldsymbol{k}_2)\cdot\boldsymbol{x}_2}
 \angbr{ \delta^\text{th}_g( \boldsymbol{k} ; \bar \eta_1 ,\boldsymbol{\hat n}_1 ) { \delta^{\text{th}\,\ast}_g}( \boldsymbol{k}' ; \bar \eta_2,\boldsymbol{\hat n}_2 ) }
 \nnn
&= \int \frac{\mathrm d^3k}{(2\pi)^3} \int_{V}\mathrm  d^3 x_1 \int_{V}\mathrm  d^3 x_2 \, 
e^{i \boldsymbol{k}\cdot(\boldsymbol{x}_1-\boldsymbol{x}_2)}
e^{-i ( \boldsymbol{k}_1\cdot\boldsymbol{x}_1
-\boldsymbol{k}_2\cdot\boldsymbol{x}_2)}  P_\text{th}  (\boldsymbol{k};\bar\eta_1, \bar\eta_2 ,\boldsymbol{\hat n}_1,\boldsymbol{\hat n}_2)\,,
\end{align}
where we have used
\beeq
 \langle \delta^\text{th}_g( \boldsymbol{k} ; \bar \eta_1 ,\boldsymbol{\hat n}_1)  \delta^{\text{th}\,*}_g( \boldsymbol{k}' ; \bar \eta_2 ,\boldsymbol{\hat n}_2) \rangle \equiv (2\pi)^3 \delta^D( \boldsymbol{k}- \boldsymbol{k}') P_\text{th}(\boldsymbol{k}; \bar\eta_1, \bar\eta_2,\boldsymbol{\hat n}_1,\boldsymbol{\hat n}_2 )\,.
\eneq
Now, we introduce the Fourier kernel $F$ that encodes the survey geometry,
\begin{equation}
F(\boldsymbol k,\boldsymbol k') \equiv   \int_{V} \mathrm d^3 x_{\rm obs} \, e^{i (\boldsymbol{k}-\boldsymbol{k}')\cdot\boldsymbol{x}_{\rm obs}}\, T(\bar \eta_z,k,\mu)   \,,
\end{equation}
where the transfer function $T$ is defined as
\beeq
P_\text{th}(\boldsymbol{k}; \bar\eta_1, \bar\eta_2,\boldsymbol{\hat n}_1,\boldsymbol{\hat n}_2 ) \equiv  P_\zeta(k)  T(\bar\eta_1,k,\mu_1)T^\ast(\bar\eta_2,k,\mu_2)\,,
\eneq
with $\mu_i$ being the cosine-angles between $\boldsymbol{k}$ and $\boldsymbol{\hat n}_i$. We can re-write the relation between the observed and theory power spectra as
\beeq
 (2\pi)^3 \delta^D (\boldsymbol{k}_1-\boldsymbol{k}_2)P_{\obs}(\boldsymbol{k_1})=\int \frac{\mathrm d^3k}{(2\pi)^3} P_\zeta(k) F(\boldsymbol k,\boldsymbol k_1)F^*(\boldsymbol k,\boldsymbol k_2)\,.
	\label{two-point_delta_obs1}
\eneq
Considering that $ (2\pi )^3\delta^D (0 ) = \int\mathrm d^3 x   \approx \int_V\mathrm  d^3 x  =V$ is approximately the observed volume, the observed power spectrum can then be derived in terms of the Fourier kernel as
\begin{equation}
P_\text{obs} (\boldsymbol{k} ) \approx \frac{1}{V}
\int \frac{\mathrm d^3k'}{ (2\pi )^3} P_\zeta ({k}' )  | F (\boldsymbol{k}', \boldsymbol{k} ) |^2
\, .
	\label{power_spectrum_obs_k'=k}
\end{equation}
In summary, the observed power spectrum is given by a convolution of
the theory power spectrum with the survey geometry represented by $\vert F (\boldsymbol{k}', \boldsymbol{k} )\vert^2 $. Throughout the paper, our primary focus will be the theory power spectrum 
in a hypersurface of simultaneity shown as red dotted in figure~\ref{light-cone}.

Since the theory Fourier mode is defined in terms of the observed 
position~$\boldsymbol{x}_{\rm obs}$, the theory power spectrum of our
interest is also defined in terms of {\it one} observed 
position~$\boldsymbol{x}_{\rm obs}$, and 
with equation~\eqref{eq:theory} it can be readily obtained by considering
the variance of the observed galaxy fluctuation as
\begin{equation}
\sigma^2_g(\boldsymbol{x}_{\rm obs})
\equiv  \langle \delta_g^{2} (\boldsymbol{x}_{\rm obs}) \rangle  =
\int \frac{\mathrm d^3k}{\left(2\pi\right)^3} \int \frac{\mathrm d^3k'}{\left(2\pi\right)^3} \langle \delta^\text{th}_g( \boldsymbol{k} ; \bar \eta_z ,\boldsymbol{\hat n} )  \delta^{\text{th}\,*}_g( \boldsymbol{k}' ; \bar \eta_z,\boldsymbol{\hat n} ) \rangle 
 \equiv \int \frac{\mathrm d^3k}{(2\pi)^3} P_\text{th}(\bm k;\bar\eta_z,\boldsymbol{\hat n})~,
 \end{equation}
 where
\begin{equation}
 P_\text{th}(\bm k;\bar\eta_z,\boldsymbol{\hat n}) = P_\zeta ({k})\left\vert T(\bar\eta_z,k,\mu)\right\vert^2\,.
\end{equation}
Given the expression of the observed galaxy fluctuation in 
equations~\eqref{ds}$-$\eqref{dnl}, we can compute the variance at a given 
observed position~$\boldsymbol{x}_{\rm obs}$ and simply read off the
theory power spectrum defined in the hypersurface set by~$\bar\eta_z$ and the observed angle $\boldsymbol{\hat n}$.
As we demonstrate in section~\ref{PS}, this theory power spectrum can be directly used to study the impact of effects such as the redshift-space distortion, but without the need of applying the distant observer approximation or other simplifying assumptions. Furthermore, since it is defined in terms of only one observed point,
there is no ambiguity involving two points such as the wide angle
effect (see, e.g.,~\cite{Yoo:2013zga}). Hereafter, we refer to $P_\text{th}(\bm k;\bar\eta_z,\boldsymbol{\hat n})$ simply as the power spectrum.

\section{Power spectrum for the observed galaxy number density}
\label{PS}

In this section, we first evaluate in section~\ref{contribution to PS} the theory power spectra of the contributions $\delta_s$, $\delta_o$ and $\delta_{nl}$, by using the method described in section~\ref{th vs obs}. The resulting expressions will be used to numerically compute the galaxy power spectrum including all general relativistic effects in section~\ref{numerical}. Then, we discuss in section~\ref{IRdiv} the issue of infrared divergences in the individual contributions, showing that they cancel out in the total galaxy power spectrum and explaining the relation to the equivalence principle. Finally, in section~\ref{RSDeL}, we consider the galaxy number density when only the standard redshift-space distortion and lensing term are accounted for.

\subsection{Individual contributions to the power spectrum}
\label{contribution to PS}

Following the method described in section~\ref{th vs obs}, the full relativistic galaxy power spectrum $P^{\text{th}}_g$ is obtained by computing the variance of the galaxy fluctuation $\delta_g$,
\begin{equation}\label{sigma}
\begin{split}
\sigma_{g}^2  (\boldsymbol{x}_{\rm obs})  &\equiv \angbr{ \delta_g^{2} (z,\boldsymbol{\hat n})  }  = \int \frac{\mathrm d^3k}{(2\pi)^3} P^{\text{th}}_g(\bm k;\bar \eta_z , \boldsymbol{\hat n}) \,.
\end{split}
\end{equation}
Since the galaxy fluctuation is given by the sum of the local and non-local contributions as $\delta_g=\delta_s+\delta_o+\delta_{nl}$, the theory Fourier mode $\delta_g^{\text{th}}(\bm k;\bar \eta_z , \boldsymbol{\hat n})$ is also the sum of the individual contributions $\delta_g^{\text{th}}(\bm k;\bar \eta_z , \boldsymbol{\hat n})=\delta_s^{\text{th}}(\bm k;\bar \eta_z , \boldsymbol{\hat n})+\delta_o^{\text{th}}(\bm k;\bar \eta_z , \boldsymbol{\hat n})+\delta_{nl}^{\text{th}}(\bm k;\bar \eta_z , \boldsymbol{\hat n})$ and
the galaxy power spectrum can be written as the sum of their power spectra 
\begin{equation}\label{Ptot}
P^{\text{th}}_g \equiv P_s + P_o + P_{nl} + 2\,P_{s\text{-}o} +2\, P_{s\text{-}nl} + 2\,P_{o\text{-}nl} \,,
\end{equation}
where the individual power spectrum is defined as
\begin{equation}
\int \frac{\mathrm d^3k}{(2\pi)^3}  P_{a\text{-}b} (\bm k;\bar \eta_z , \boldsymbol{\hat n}) = \frac{1}{2} \sbr{ \langle \delta_a^{}(z , \boldsymbol{\hat n})\delta_b^*(z , \boldsymbol{\hat n})\rangle + \angbr{ \delta_b^{}(z , \boldsymbol{\hat n}) \delta_a^* (z , \boldsymbol{\hat n})} }\,,
\end{equation}
and $P_a \equiv P_{a\text{-}a}$.
All the power spectra are evaluated given the observed redshift $z$ and observed angle $\boldsymbol{\hat n}$.
Note that the splitting of $P^{\text{th}}_{g}$ as in equation~\eqref{Ptot} is gauge-dependent, as it simply follows from the decomposition of $\delta_g$ into the local and non-local contributions. However, such decomposition is convenient to compare the complete prediction $P_{g}^{\texth}$ with the previous work in literature. For example, the power spectrum computed in \cite{Yoo:2010,Yoo:2012,Jeong} corresponds to $P_s$ alone, as we show below in detail.

First, we compute the power spectrum $P_s$ of the fluctuation $\delta_s$ at the source position and compare it with the result presented in \cite{Yoo:2010,Jeong,Yoo:2012}.
The variance of $\delta_s$ can be directly derived starting from equation~\eqref{ds} as
\begin{equation}
\begin{split}
\sigma_s^2  (\boldsymbol{x}_{\rm obs})  = \angbr{ \delta^2_s (z,\boldsymbol{\hat n}) }
=  \int \frac{\mathrm d^3 k}{(2\pi)^3}  \bigg[&  \big( b^2  + 2 bf  \mu_k^2 + f^2    \mu_k^4 \big)   D^2
\\
&
+ \big( 2bD\mathcal A + 2Df \mathcal A \,  \mu_k^2 + \mathcal B^2  \mu_k^2 \big)  \frac{1}{k^2}  +  \mathcal A^2  \frac{1}{ k^4}     \bigg] P_m(k)  \,,
\end{split}
\end{equation}
where $\mu_k = \boldsymbol{\hat k}\cdot\boldsymbol{\hat n}\,$ and $P_m(k)$ is the matter power spectrum at redshift zero.
The corresponding power spectrum can now be read off as
\begin{equation}
P_s(\bm k; \bar \eta_z,\boldsymbol{\hat n}) =  P_z(\bm k; \bar \eta_z,\boldsymbol{\hat n}) +    \bigg[    \big(  2bD\mathcal A + 2 Df\mathcal A  \mu_k^2 + \mathcal B^2  \mu_k^2 \big)  \frac{1}{ k^2} +  \mathcal A^2  \frac{1}{ k^4}
    \bigg] P_m(k)  \,,
\end{equation}
where we introduced the standard redshift-space power spectrum \cite{kaiser},
\begin{equation}
P_z(\bm k; \bar \eta_z,\boldsymbol{\hat n})\equiv \big[ b^2  + 2 b  f  \mu_k^2 + f^2 \mu_k^4   \big]D^2 P_m(k) \,.
\end{equation}

The power spectrum $P_s$ at the source position recovers the standard redshift-space power spectrum $P_z$ without the need to assume the distant-observer approximation, demonstrating the utility of our method.
 Our theory power spectrum is the one defined in an (infinite) hypersurface at the observed redshift $z$ that provides the correct observed variance $\sigma^2$ at $\bm x_{\rm obs}$. Since it does not involve the survey geometry or two observed positions it is well suited for theoretical investigations.
Furthermore, this result $P_s$ is in agreement with~\cite{Yoo:2010, Jeong,Yoo:2012}. The contribution at the source to the power spectrum consists of the redshift-space power spectrum $P_z$ and the general relativistic effects, corresponding to the terms in the square brackets that are proportional to $k^{-2}P_m(k)$ and $k^{-4}P_m(k)$. 
Interestingly, since $P_m(k) \propto k^{n_s}$ at small $k$, with the spectral index $n_s \approx 0.96$, these relativistic corrections in $P_s$ diverge when $k$ goes to zero. As a result, $P_s$ is infrared-divergent (so is $\sigma^2_s$ from $\delta_s$ alone), as already pointed out in~\cite{Yoo:2010,Jeong,Yoo:2012}. As we will discuss in section~\ref{IRdiv}, such infrared divergence is not physical, as the source contribution $P_s$ alone is not an observable and the equivalence principle is violated in its expression. 
Only the expression for the total power spectrum $P^{\text{th}}_g$ is consistent with the equivalence principle. 

As we have already mentioned, the splitting of $P^{\text{th}}_g$ into individual components is for convenience, and these components are not by themselves measurable physical quantities. Nevertheless, considering the contributions individually can be useful to understand their importance on different scales for $P^{\text{th}}_g$. With the expression for $P_s$ at hand, we can now derive the remaining contributions which we expect to contain terms proportional to $k^{-2}P_m(k)$ and $k^{-4}P_m(k)$ that eventually cancel those in $P_s$ when summed all together. In addition, there can be terms proportional to $P_m(k)$ (non-divergent terms) that would result in deviations from the standard prediction $P_z$ of the redshift-space distortion. As we will show, such deviations occur on large scales and they are redshift-dependent.

Using the same method applied to obtain $P_s$, we derive the power spectrum of the non-local contributions,
\begin{align}
P_{nl}( \bm k) = \int_0^{\bar r_z} \mathrm d\bar r_{1}\int_0^{\bar r_z} \mathrm  d\bar r_{2} \, e^{i k \Delta r \mu_k}
&\bigg[ \mathcal E_1 \mathcal E_2 \frac{1}{k^4}-2\mathcal E_1\mathcal F_2\frac{i\mu_k}{k^3}   + 2 \mathcal E_1\mathcal G_2 \frac{ (1- \mu_k^2)}{k^2}+ \mathcal F_1 \mathcal F_2 \frac{ \mu_k^2 }{k^2}  \nnn
&\quad -2\mathcal G_1\mathcal F_2(1-\mu_k^2)\frac{i\mu_k}{k}+ \mathcal G_1\mathcal G_2 \big( 1-\mu_k^2 \big)^2\bigg] P_m(k) \,,
\end{align}
and the power spectrum at the observer position,
\begin{align}
P_o(\bm k) &= \bigg[ \mathcal C^2 \frac{1}{k^4}  +  \mathcal D^2 \frac{ \mu_k^2  }{k^2}  \bigg] P_m(k) \,, 
\end{align}
where we omitted the dependence on observed redshift and angle in our notation, $P_{nl}(\mathbf k)\equiv P_{nl}(\mathbf k;\bar\eta_z,\boldsymbol{\hat n})$ and $P_{o}(\mathbf k)\equiv P_{o}(\mathbf k;\bar\eta_z,\boldsymbol{\hat n})$, and we defined $\Delta r \equiv \bar r_1 - \bar r_2$ and $\mathcal X_i \equiv \mathcal X(\bar r_i)$ for $\mathcal X \equiv \mathcal E,\,\mathcal F ,\,\mathcal G$.

The above expressions can be expanded into angular multipoles in terms of the Legendre polynomials $L_\ell(\mu_k)$ as $P(\boldsymbol{k}) \equiv \sum_{\ell=0}^{\infty} L_\ell(\mu_k) P_\ell(z,k) $.
We obtain
\begin{align}
\label{Ps}
P_s( \bm k) &=  P_z(\bm k) +\sbr{  L_0   \frac{\mathcal A^2 }{ k^4}  + \frac{1}{k^2}\bigg({2 D\mathcal A}\Big(b+\frac{f}{3}\Big) +\frac{\mathcal B^2}{3} \bigg) L_0  + \frac{2}{3k^2} (2 Df\mathcal A  + \mathcal B^2)L_2 } P_m(k)    \,,
\\\label{Pz}
P_z(\bm  k) &=  \bigg[\bigg( b^2    + \frac{2}{3} b f
    + \frac{1}{5} f^2   \bigg) L_0   +  \bigg( \frac43  b f  + \frac4 7 f^2   \bigg) L_2  + \frac{8}{35} f^2 L_4   \bigg] D^2 P_m(k)  \,,
\end{align}
for the expanded source and standard redshift-space power spectrum, where we omitted the dependence of the Legendre polynomials on $\mu_k$ in our notation, $L_l\equiv L_l(\mu_k)$. For the expanded observer power spectrum, we obtain
\begin{align}
P_o( \bm k) &= \bigg[ \mathcal C^2 L_0 \frac{1}{k^4}  +\frac{1}{3}  \mathcal D^2 (L_0 +2L_2)  \frac{1}{k^2}  \bigg] P_m(k)   \,,\label{Po}
\end{align}
and, for the expanded non-local power spectrum, 
\begin{align}
\label{Pnl}
P_{nl}(\bm k)=&  \sum_{n=0}^\infty (-1)^n (4n+1)L_{2n} \int_0^{\bar r_z} \mathrm d\bar r_{1}\int_0^{\bar r_z}\mathrm  d\bar r_{2} \,
\bigg[ \mathcal E_1 \mathcal E_2 \frac{1}{k^4} j_{2n}(\Delta x ) -2\mathcal E_1\mathcal F_2\frac{j'_{2n}(\Delta x)}{k^3}
\nnn
& \quad
 - \mathcal F_1\mathcal F_2 \frac{1 }{k^2}  j_{2n}''(\Delta x )+ 2 \mathcal E_1\mathcal G_2 \frac{1}{k^2}\rbr{ j_{2n}(\Delta x ) + j_{2n}''(\Delta x )}-2\mathcal G_1\mathcal F_2\frac{1}{k}\rbr{j'_{2n}(x)+j'''_{2n}(x)}
\nnn
&\quad + \mathcal G_1\mathcal G_2\rbr{  j_{2n}(\Delta x )  + 2 j_{2n}''(\Delta x ) + j_{2n}''''(\Delta x )}  
\bigg] P_m(k) \,,
\end{align}
where $j_{\ell}(y)$ are the spherical Bessel functions, $j_{\ell}'(y)\equiv\partial_y j_\ell(y)$ are their derivatives, and we defined $\Delta x\equiv k \Delta r$.
Note that we used the relations
\begin{align}
& \mu_k^2=\frac{1}{3}[L_0(\mu_k)+2L_2(\mu_k)] \,, \qquad \mu_k^4=\frac{1}{35}[7 L_0(\mu_k) +20 L_2(\mu_k)+ 8 L_4(\mu_k)]\,,
\end{align}
and the plane wave expansion 
\begin{align}
& e^{i y \mu_k} = \sum_{\ell=0}^{\infty} i^\ell (2\ell+1) j_\ell(y) L_\ell(\mu_k)\,, \label{planewave}
\end{align}
along with the derivatives of this equation up to 4th order. 
The cross power spectra $P_{s\text{-}o}$, $P_{s\text{-}nl}$, $P_{o\text{-}nl}$ of $\delta_s$, $\delta_o$ and $\delta_{nl}$ are derived in the same way and their expressions are presented in appendix~\ref{A}. 
The standard redshift-space power spectrum in equation~\eqref{Pz} shows the usual decomposition into the monopole, quadrupole and hexadecapole. However, the non-local contribution from the line-of-sight integration gives rise to the contributions at all even multipoles $\ell = 0,2,4,6,8,\dots$ (odd multipoles vanish, which can be seen from the fact that $j_n(-\Delta x)=(-1)^nj_n(\Delta x)$ and the fact that $\bar r_1$ and $\bar r_2$ can be interchanged in the integral). 
The results in equations~\eqref{Ps}--\eqref{Pnl} and \eqref{Pso}--\eqref{Ponl} provide the complete analytical expression for the fully relativistic power spectrum of the observed galaxy fluctuation.

\subsection{Infrared divergences and their cancellation}\label{IRdiv}

In previous work on general relativistic effects in galaxy clustering (see, e.g., \cite{Yoo:2010, Jeong,Yoo:2012}), it was concluded that they produce infrared-divergent terms in the galaxy power spectrum. Indeed, the expression for the source power spectrum given in equation~\eqref{Ps} contains divergent terms $\propto k^{-2}P_m(k)$ and $\propto k^{-4}P_m(k)$,
\begin{equation}\label{pk0s}
P_s^{\,\text{div.}}(\bm k)= \bigg[  \mathcal A^2 L_0   \frac{1}{ k^4} +   \bigg( 2bD\mathcal A + \frac{2}{3} Df \mathcal A  +\frac{1}{3} \mathcal B^2\bigg) L_0\frac{1}{ k^2}  + \frac{2}{3} (2 Df \mathcal A  + \mathcal B^2)L_2  \frac{1}{ k^2}     
     \bigg] P_m( k)  \,,
\end{equation}
where $P_s^{\,\text{div.}}(\bm k)=P_s(\bm k)$ for very long modes $k\to0$. Such divergent terms have previously been considered as a major contamination to the signal of primordial non-Gaussianity which is also believed to produce terms $\propto k^{-2}$ (see, e.g., \cite{Jeong, nG1, nG2, nG3, nG4, nG5, nG6}). However, the quantity $\delta_s$ is, by itself, not observable. Only the total galaxy number density $\delta_g$, given by the sum $\delta_g=\delta_o+\delta_{nl}+\delta_s$ of all terms evaluated at the source and observer position and along the line of sight, is an observable quantity. To investigate whether the total galaxy power spectrum $P_g(\bm k)$ has any infrared-divergence, we need to sum up the divergent terms in all individual power spectra and cross-power spectra,
\beeq
P_g^{\,\text{div.}}(\bm k)=P^{\,\text{div.}}_s(\bm k) + P^{\,\text{div.}}_o(\bm k) + P^{\,\text{div.}}_{nl}(\bm k) + 2\,P^{\,\text{div.}}_{s\text{-}o}(\bm k) +2\, P^{\,\text{div.}}_{s\text{-}nl}(\bm k) + 2\,P^{\,\text{div.}}_{o\text{-}nl}(\bm k) \,,
\eneq
where
\begin{align}
P_o^{\,\text{div.}}(\bm k) &=  \bigg[ \mathcal C^2 L_0 \frac{1}{k^4}  +\frac{1}{3}  \mathcal D^2 \rbr{L_0 +2 L_2}  \frac{1}{k^2}  \bigg] P_m(k)  \,, \label{Podiv}
\end{align}
and
\begin{align}
P_{nl}^{\,\text{div.}}(\bm k) = \int_0^{\bar r_z}\mathrm d\bar r_{1}\int_0^{\bar r_z}\mathrm d\bar r_{2} \,&
\bigg[  \mathcal E_1 \mathcal E_2  \rbr{ \rbr{ 1  - \frac 16  \Delta x ^2 }L_0 
- \frac 13 \Delta x^2  L_2}\frac{1}{k^4}+\frac{2}{3}\mathcal E_1\mathcal F_2\frac{\Delta r}{k^2}\rbr{L_0+2L_2}
\nnn
&\quad
 + \frac 43 \mathcal E_1\mathcal G_2  \rbr{ L_0 - L_2 }\frac{1}{k^2}+ \frac 13 \mathcal F_1 \mathcal F_2   \rbr{ L_0  +2 L_2 }\frac{1 }{k^2} \bigg] P_m(k) \,. \label{pk0nl}
\end{align}
The divergent parts of the cross power spectra of $\delta_s$, $\delta_o$ and $\delta_{nl}$ are given in equations~\eqref{Pk0so}$-$\eqref{Pk0onl}. Note that to obtain the divergent contributions to the power spectra in equations~\eqref{Pnl} and \eqref{Pso}$-$\eqref{Ponl}, we expanded the Bessel functions around zero up to the order required to have terms scaling as $k^{-4}P_m(k)$ and $k^{-2}P_m(k)$ in the power spectra. Furthermore, note that according to the multipole expansion, the divergent contributions only appear in the monopoles and quadrupoles of the power spectra, i.e.~the quantities proportional to $L_0$ and $L_2$. Now, to obtain the total $P_g^{\,\text{div.}}(\bm k)$, we use that 
\beeq
\mathcal A+\mathcal C+\int_0^\rz\dr\mathcal E(\bar r)=0\,,\qquad \mathcal A\rz+\mathcal B+\mathcal D+\int_0^\rz\dr\mathcal F(\bar r)+\int_0^\rz\dr\mathcal E(\bar r)\bar r=0\,, \label{z0vanish}
\eneq
which can be shown by applying the relations between the growth functions $D_\Psi$ and $D_V$ given in equation~\eqref{relgf}. While this equation might seem somewhat arbitrary at this point, we explain below that it indeed has a physical interpretation, as it ensures that the uniform potential and uniform gravitational force have no contribution to the observable $\delta_g$. Summing up all the respective contributions from equations~\eqref{pk0s}, \eqref{Podiv}, \eqref{pk0nl} and \eqref{Pk0so}--\eqref{Pk0onl}, we see that the resulting expression for $P_g^{\,\text{div.}}(\bm k)$ can be fully factorized using these vanishing expressions. Hence, while all the individual contributions, which are by themselves not measurable, exhibit a diverging behavior at very large scales, this infrared divergence vanishes completely in the total galaxy power spectrum, $P_g^{\,\text{div.}}(\bm k)=0$.

Having obtained this result, one might wonder whether there is a physical meaning behind the fact that the sum of all divergent terms turns out to be exactly zero. Indeed, this is not a random coincidence, but a manifestation of the equivalence principle. This fundamental principle of general relativity offers, along with the gauge invariance, a unique way to test the validity of theoretical predictions for observable quantities, such as the observed galaxy fluctuation $\delta_g$ and its power spectrum. It was shown in~\cite{Yoo,Fulvio} that the expression for the galaxy fluctuation in equations~\eqref{dg}--\eqref{nl} indeed fulfills these two requirements. Furthermore, it was demonstrated in~\cite{SG divergence,YOGO15} that the compatibility with the equivalence principle ensures that the expression does not exhibit any infrared divergence on super horizon-scales. Now, we revisit this issue for the galaxy power spectrum, clarifying the relation between the equivalence principle and the exact vanishing of $P_g^{\,\text{div.}}(\bm k)$.

The equivalence principle states that, as a consequence of the equality of inertial and gravitational mass, an observer cannot distinguish the acceleration caused by a uniform gravitational field from the acceleration caused by a non-inertial reference system accelerating itself. 
This implies that a uniform gravitational field cannot have any effect on physical observables such as the galaxy number density $\delta_g$. Note that the requirement of uniformity of the gravitational field means that over the scale of the experiment there cannot be any tidal forces that would induce a measurable tidal motion between two test particles. Considering a survey observing galaxies at some redshift $z$, the sources and the observer are located on the past light-cone separated by the (comoving) distance $\bar r_z$. Hence, any perturbation behaving as a uniform gravitational field on the scale set by the redshift $z$, acting equally on all test particles (i.e., the galaxies), should not affect the observable galaxy number density. 

To extract the effect of the uniform gravitational field from the expression for the galaxy number density, note that the potential
\begin{equation}
\Psi(\eta,\bar r  \boldsymbol{\hat n}) =\int \frac{\mathrm d^3 k}{(2\pi)^3} e^{i \boldsymbol{k}\cdot \bar r \boldsymbol{\hat n}} \Psi(\boldsymbol{k};\eta)\,,
\end{equation}
can be, by expanding the exponential in terms of $k \bar r$, written as  
\begin{equation}
	\Psi = \Psi_o + \bar r \, \Psi_1 +  \sum_{n \geq 2} \frac{\bar r^{n}}{n!}   \Psi_n \,,
\end{equation}
where
\begin{equation}
	\Psi_o   \equiv \int  \frac{\mathrm d^3 k}{(2\pi)^3} \Psi(\eta,\boldsymbol{k}) \,, \qquad \Psi_1 \equiv \hat n^i  \big[ \partial_i \Psi  \big]_o \equiv \hat n^i \int \frac{\mathrm d^3 k}{(2\pi)^3}\, i k_i  \Psi(\eta,\boldsymbol{k})\,,  \qquad \Psi_n \equiv \partial_{\parallel}^n \Psi \big\vert_{\bar r=0} \,.
\end{equation}
The first two terms in the expansion, $\Psi_o$ and $\Psi_1$, i.e.~the uniform gravitational potential and the uniform gravitational force, do not produce any tidal forces and thus cannot have any effect on physical observables measured by the observer. Note that, the contribution of a given $k$-mode to $\Psi_n$ is proportional to $(k\bar r)^n$. Hence, any $k$-mode beyond the scale of the survey, i.e.~with $k\bar r_z\ll 1$, falls off fast with increasing $n$. This means that the equivalence principle implies in particular that such very long modes do not affect the observable.

Now, by considering the variances $\sigma_o^2$ and $\sigma_1^2$ of $\Psi_o$ and $\Psi_1$,
\begin{equation}\label{sigmao1}
\sigma_o^2=\langle \Psi_o \Psi_o \rangle  \propto \int  \frac{\mathrm d^3 k}{(2\pi)^3} \frac{1}{k^4} P_m(k) \,,
\qquad
\sigma_1^2=\langle \Psi_1 \Psi_1 \rangle  \propto \int \frac{\mathrm d^3 k}{(2\pi)^3}\frac{\mu_k^2 }{k^2}  P_m(k) \,,
\end{equation}
we immediately see that the presence of such terms would yield terms $\propto k^{-4}P_m$ and $\propto k^{-2}P_m$, hence leading to an infrared-divergence in the power spectrum of the observable. However, it was shown in~\cite{Fulvio} that the full gauge-invariant expression of the galaxy fluctuation $\delta_g$ is compatible with the equivalence principle, i.e.~does not contain any contributions from $\Psi_o$ and $\Psi_1$. Hence, considering this fundamental principle we indeed would not expect any divergent terms in the galaxy power spectrum, in full agreement to the above result $P_g^{\,\text{div.}}(\bm k)=0$.

 To further illustrate the correspondence between the equivalence principle and the previously proven vanishing of the infrared-divergence in the power spectrum, we revisit the proof given in~\cite{Fulvio}, using our decomposition of $\delta_g$ into observer terms $\delta_o$, source terms $\delta_s$ and non-local terms $\delta_{nl}$. The gravitational potential is related to the curvature potential as
\beeq
\Psi(\eta,\bar r\mathbf{\hat n})=D_\Psi(\eta)\zeta(\bar r\mathbf{\hat n})=D_\Psi(\eta)\sbr{\zeta_o+\bar r\zeta_1(\mathbf{\hat n})+\dots}\,,
\eneq
which implies
\beeq
v(\eta,\bar r\mathbf{\hat n})=-D_V(\eta)\sbr{\zeta_o+\bar r\zeta_1(\mathbf{\hat n})+\dots}\,,\qquad V_\parallel(\eta,\bar r\mathbf{\hat n})=D_V(\eta)\sbr{\zeta_1+\dots}\,,
\eneq
for the long-mode velocity potential and line-of-sight velocity. Inserting this into equations~\eqref{dg}--\eqref{nl}, we see that the contributions of $\zeta_o$ and $\zeta_1$ to $\delta_o$, $\delta_s$ and $\delta_{nl}$ are given by
\begin{align}
\delta_s(\zeta_o,\zeta_1)&=-\frac{1}{C}\sbr{\mathcal A\zeta_o+\rbr{\mathcal A\rz+\mathcal B}\zeta_1}\,,\\
\delta_o(\zeta_o,\zeta_1)&=-\frac{1}{C}\sbr{\mathcal C\zeta_o+\mathcal D\zeta_1}\,,\\
\delta_{nl}(\zeta_o,\zeta_1)&=-\frac{1}{C}\int_0^\rz\dr\,\sbr{\mathcal E(\bar r)\zeta_o+\rbr{\mathcal E(\bar r)\bar r+\mathcal F(\bar r)}\zeta_1}\,.
\end{align}
Summing up these equations, we see that the total contributions of $\zeta_o$ and $\zeta_1$ to $\delta_g$ are indeed exactly vanishing according to equation~\eqref{z0vanish}, which is then used to obtain the result $P_g^{\,\text{div.}}(\bm k)$. 

We conclude that the diverging power in the galaxy power spectrum at very low~$k$ found in previous work is originating from 
the consideration of only~$P_s$, i.e.~the relativistic effects at the source position,
while ignoring other relativistic contributions in the full gauge-invariant
expression. With all relativistic effects accounted for in the power
spectrum, we showed that there exist {\it no} relativistic contributions 
to the power spectrum that scale with $k^{-2}$ or $k^{-4}$ at low~$k$, which is indeed the result expected from the equivalence principle. Note that to obtain this result, we have treated the potential at the observer position as a statistical field instead of a fixed value. Indeed, when taking into account non-local contributions, treating fields at all $\bar r>0$ statistically, but not at $\bar r=0$ (see~\cite{Vincent}), would be an obscure discontinuity. Nevertheless, taking statistical averages at the observer position might seem spurious to some readers, as we can only take observations from one fixed position. This issue was closely investigated in~\cite{ObserverTerms}, where it was concluded that the error from having only one observer position is already comprised in the cosmic variance. This is the fundamental error arising from observing only one realization of the universe and, importantly, from only one vantage point. Hence, except from cosmic variance, treating the fields at the observer position statistically does not lead to any mistake or inconsistency in the measurement. 

Finally, note that while there are no divergent terms in the galaxy power spectrum, mimicking the signature of primordial non-Gaussianity, other effects (represented by non-divergent terms) manifest on large scales (but not in the limit $k\to 0$). Hence, despite the vanishing of the infrared divergence, the standard redshift-space power spectrum does not accurately describe the large-scale behavior of the galaxy power spectrum. We discuss this in detail in section~\ref{numerical}.

\subsection{Redshift-space distortion and gravitational lensing}\label{RSDeL}

In this section, we consider the galaxy number density taking only the standard redshift-space distortion given in equation~\eqref{deltaz} and the standard gravitational lensing given in equation~\eqref{deltaL} into account,
\begin{equation}
	\delta_{ st}  = \delta_z + \delta_L \,,
\end{equation}
where we refer to this equation as the standard expression for the galaxy number density. While the redshift-space distortion power spectrum is well studied, the gravitational lensing
contribution to the power spectrum has found little consideration in theoretical work (see, however, \cite{Hui:2007tm}).
Though the gravitational lensing contribution changes across the sky,
it changes little when the source position is moved along the line-of-sight
within the survey volume, since the lensing contribution arises from the
matter density fluctuations around half the distance to the survey volume
and the survey depth is typically smaller than the distance to the survey.
Consequently, the gravitational lensing contribution to the power spectrum
is very close to the Dirac delta function along the line-of-sight 
wavevector~$k_\parallel$,
while it is similar to the angular power spectrum along the transverse
wavevector~$k_\perp$. In reality, its contribution to the power spectrum differs from the
Dirac delta function in the ideal case, rather its shape is largely determined by the survey
window function (see, e.g., \cite{Hui:2007tm}). 
In contrast, our approach to the theory power spectrum can exactly
compute the gravitational lensing contribution to the power spectrum,
independent of survey geometry, which is another advantage of our method.

To evaluate the standard power spectrum $P_{st}(\boldsymbol{k})$ of the standard galaxy number density $\delta_{st}$, first note that equation~\eqref{deltaz} for $\delta_z$ can be written as
\begin{equation}
	\delta_{z} =\int \frac{\mathrm d^3 k}{(2\pi)^3}   e^{i k \bar r_z\mu_k} \rbr{ b   +  f   \mu_k^2 } D     \delta(\bm k) \,,
\end{equation}
and equation~\eqref{deltaL} for $\delta_L$ as
\begin{equation}
	\delta_{L}  = \int \frac{\mathrm d^3 k}{(2\pi)^3}     \int_0^{\bar r_z}\mathrm d\bar r\, e^{i k \bar r \mu_k} \sbr{\mathcal F(\bar r)\frac{i\mu_k}{k}+\mathcal G(\bar r)(1-\mu_k^2)}   \delta(\bm k)\,.
\end{equation}
The power spectrum $P_{st}(\boldsymbol{k})$ is given by 
\begin{equation}
	P_{st}(\boldsymbol{k})= P_{z}(\boldsymbol{k})+ P_L(\boldsymbol{k}) + 2\, P_{z\text{-}L}(\boldsymbol{k})  \,,
\end{equation}
where the first term is the standard redshift-space power spectrum given in equation~\eqref{Pz},
the second term is the lensing power spectrum obtained from $\langle\delta_{L}^{2}(z,\boldsymbol{\hat n})\rangle$:
\begin{align}
P_{L}(\bm k)=&  \sum_{n=0}^\infty (-1)^n (4n+1)L_{2n} \int_0^{\bar r_z} \mathrm d\bar r_{1}\int_0^{\bar r_z}\mathrm  d\bar r_{2} \,
\bigg[ - \mathcal F_1\mathcal F_2 \frac{1 }{k^2} j_{2n}''(\Delta x ) -2\mathcal G_1\mathcal F_2\frac{1}{k}\rbr{j'_{2n}(x)+j'''_{2n}(x)}
\nnn
&\quad + \mathcal G_1\mathcal G_2\rbr{  j_{2n}(\Delta x )  + 2 j_{2n}''(\Delta x ) + j_{2n}''''(\Delta x )}  \bigg] P_m(k) \,,\label{PL}
\end{align}
and the last term is the cross power spectrum obtained  from $\langle\delta_{z}^{}(z,\boldsymbol{\hat n})\delta_{L}(z,\boldsymbol{\hat n}) \rangle$,
\begin{align}
 P_{z\text{-}L}(\bm k) =  &\int_0^{\bar r_z}\mathrm d\bar r\  \sum_{n=0}^\infty (-1)^n (4n+1)L_{2n} 
 \bigg[	
Df \mathcal F(\bar r) \frac{1}{k}  j_{2n}'''(\Delta x_z) - b D \mathcal F(\bar r) \frac{1}{k}   j_{2n}'(\Delta x_z)
\nnn
& + b D   \mathcal G(\bar r)  \rbr{ j_{2n}(\Delta x_z) +  j_{2n}''(\Delta x_z) }
 - Df\mathcal G(\bar r)  \rbr{ j_{2n}''(\Delta x_z) + j_{2n}''''(\Delta x_z)} \bigg]
P_m(k)    \,,\label{PK-L}
\end{align}
with $\Delta x_z \equiv k (\bar r_z-\bar r)$. 
Note that equations~\eqref{PL} and~\eqref{PK-L} can be obtained from equations~\eqref{Pnl} and \eqref{Psnl} by setting $\mathcal A \equiv \mathcal B \equiv \mathcal E \equiv 0\,$, as the contributions proportional to these functions do not appear in~$\delta_{st}$.

While, as mentioned above, our approach allows us to compute the contribution of the standard lensing independently of the survey volume, one problem becomes evident when considering equation~\eqref{PL}: It contains a term $\propto k^{-2} P_m(k)$, which leads to an infrared divergence. As we have argued in section~\ref{IRdiv}, this divergence is an unphysical artifact of not considering all relativistic effects consistently. One possible way of computing the contribution of the standard lensing alone would be to introduce an ``IR-safe'' lensing power spectrum,
\begin{align}
P^{\text{IR-safe}}_{L}(\bm k)\equiv&  \sum_{n=0}^\infty (-1)^n (4n+1)L_{2n} \int_0^{\bar r_z} \mathrm d\bar r_{1}\int_0^{\bar r_z}\mathrm  d\bar r_{2} \,
\bigg[ - \mathcal F_1\mathcal F_2 \frac{1 }{k^2}\rbr{j_{2n}''(\Delta x ) -\frac 13 \delta_{n0}-\frac{2}{15}\delta_{n1}} \nnn
& -2\mathcal G_1\mathcal F_2\frac{1}{k}\rbr{j'_{2n}(x)+j'''_{2n}(x)} + \mathcal G_1\mathcal G_2\rbr{  j_{2n}(\Delta x )  + 2 j_{2n}''(\Delta x ) + j_{2n}''''(\Delta x )}  \bigg] P_m(k) \,,
\end{align}
removing the divergent contribution $k^{-2}P_m(k)$. However, the presence of
the infrared divergence indicates that the standard gravitational lensing
alone is, in fact, not observable. Therefore, we will compute the full relativistic 
contributions to the power spectrum, decomposed in terms of angular multipoles, in section~\ref{numerical}.

\section{Numerical computation of the power spectrum}\label{numerical}

In this section, we numerically investigate the theory power spectrum and its different contributions.  
For numerical calculations we assume a flat $\Lambda$CDM universe with dark matter density $\Omega_{cdm}h^2=0.12$, baryon density $\Omega_bh^2 = 0.0462$, scalar amplitude $A_s=2.1 \times 10^{-9}$ at the pivot scale $k_0 = 0.05 \,\text{Mpc}^{-1}$, spectral index $n_s=0.966$ and Hubble parameter $h=0.674$, consistent with the Planck 2018 results~\cite{Planck2018}. Furthermore, we assume $b=1$, no magnification bias and set the evolution bias to $e=3$ at any redshift. 
Using \texttt{CLASS}~\cite{CLASS}, we obtain the matter power spectrum $P_m( k)$ today, and then we use the relations in section~\ref{metric and solutions} to obtain the power spectra of the other scalar perturbations at $a_o=1$. With the respective growth functions also given in section~\ref{metric and solutions}, we are fully equipped to numerically evaluate the resulting power spectra at a general scale factor $a$, and compute their contributions to the total galaxy power spectrum.

In the next three subsections, we compute the total monopole ($\ell=0$), quadrupole  ($\ell=2$) and hexadecapole  ($\ell=4$), taking into account the complete gauge-invariant expression of the galaxy number density in equations~\eqref{dg}--\eqref{nl}, i.e.~accounting for all the contributions at the source, at the observer and along the line of sight. Higher order multipoles $\ell\ge 6$ have little significance. The standard redshift-space power spectrum, which is the dominant contribution on small scales, does not contribute to any multipoles beyond $\ell=4$, which means that there is no significant contribution to such multipoles for large $k$. For low $k$, the contribution to the multipoles is decreasing with larger $\ell$ due to the behavior of the spherical Bessel functions $j_\ell(x)$ at low $x$ ($x\ll l$): The position of the first peak is increasing with its height decreasing for larger $\ell$, and $j_\ell(x)$ is remaining close to zero for an increasingly large interval. For these reasons, we consider only the monopole, quadrupole and hexadecapole in our numerical evaluations, and compare the total power spectra with their standard redshift-space predictions. Note that, to deal with the derivatives of the spherical Bessel functions in the equations for the multipole power spectra, we use the relation
\begin{equation}\label{bessel}
j'_{n}(y) = \frac{1}{2n+1}\rbr{n j_{n-1}(y) -(n+1) j_{n+1}(y)} \,,
\end{equation}
to avoid computing the derivatives at every step in the integrations.

\begin{figure}
\begin{center}\includegraphics[width=12cm]{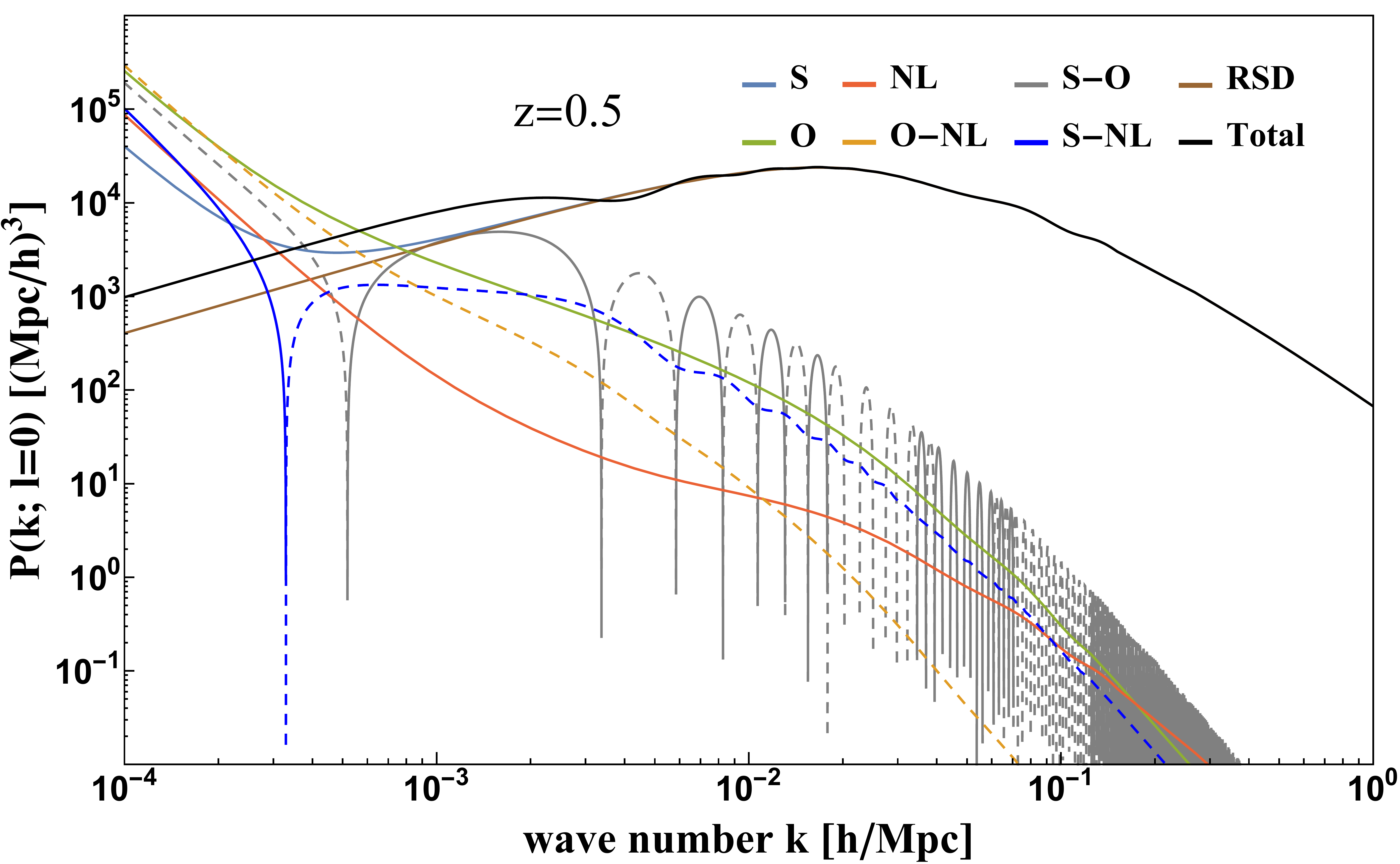}\end{center}
\caption{Monopole power spectrum at redshift $z=0.5$ as a function of $k$, including the total power spectrum $P_g$ (black line), the standard redshift-space power spectrum $P_z$ (brown line), and the individual contributions of the power spectra and cross power spectra of the source, observer and non-local terms (other colors). Here and in all other plots of this section dashed lines represent negative values. Note that the individual contributions exhibit an IR-divergent behavior, which vanishes for the total monopole power spectrum. Furthermore, note that, as stated in equation~\eqref{Ptot}, the cross power spectra contribute to the total with a factor of 2. This factor is accounted for in the curves for the cross power spectra, in this and all other plots of section~\ref{numerical}.} \label{P0z05div}
\end{figure}

\subsection{Monopole} \label{monopole}

First, we numerically evaluate the contributions to the monopole power spectrum, given by the terms proportional to $L_0$ in equations~\eqref{Ps}--\eqref{Pnl} and~\eqref{Pso}--\eqref{Ponl}, and show the resulting curves for redshift $z=0.5$ in figure~\ref{P0z05div}. This figure illustrates the divergent behavior of the individual components in the infrared, in particular of the source power spectrum $P_s$, which was taken seriously in theoretical interpretations. However, as illustrated in figure~\ref{P0z05div} and proved analytically in section~\ref{IRdiv}, this divergent behavior vanishes for the total monopole galaxy power spectrum $P_g(k; l=0)$, yielding no corrections to the signal of primordial non-Gaussianity. Nevertheless, we can see additional features in the total monopole power spectrum $P_g(k; l=0)$ (black curve) compared to the standard redshift-space prediction $P_z(k;l=0)$ (brown curve) at scales $k\lesssim k_{eq}$, where $k_{eq}$ corresponds to the scale of matter-radiation equality at which the turn-over in the power spectrum occurs.
\begin{figure}\begin{center}
\includegraphics[width=12cm]{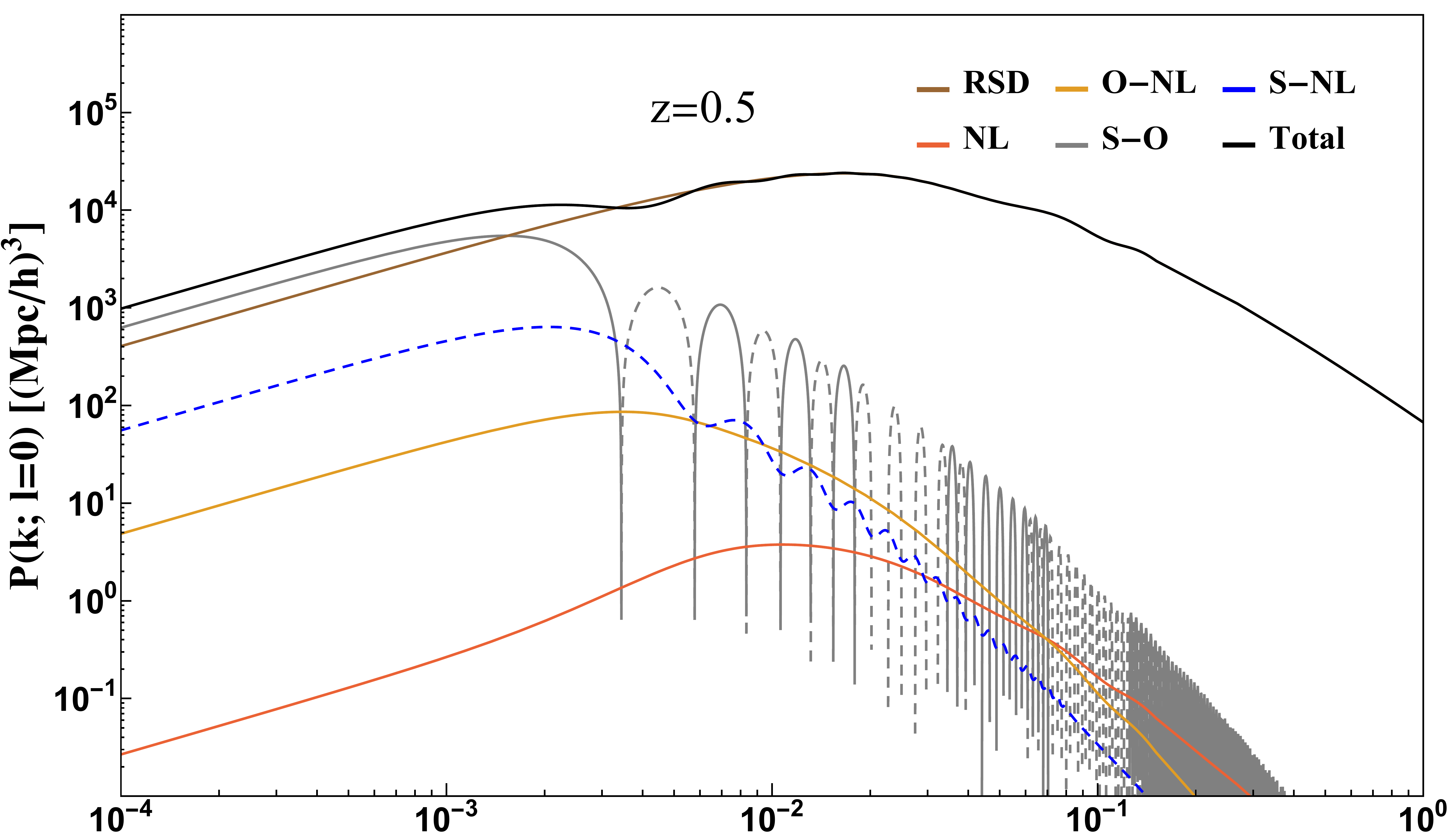} 
\includegraphics[width=12cm]{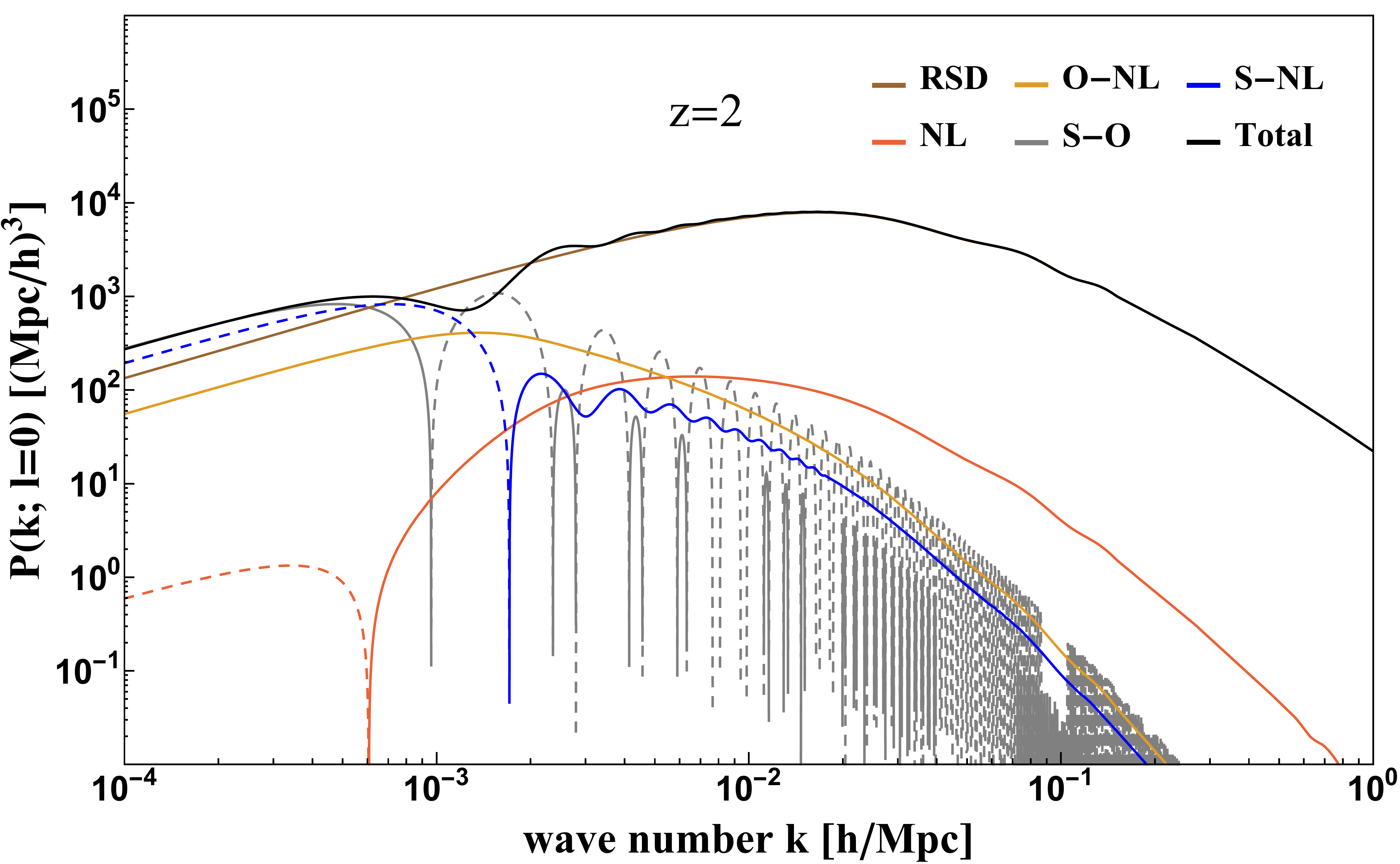}
\caption{The total monopole power spectrum and the individual, IR-safe contributions at redshift $z=0.5$ (top) and $z=2$ (bottom). Note that there are no curves for the contributions of the IR-safe source and observer power spectra, as the former is identical to the standard redshift-space power spectrum and the latter is fully vanishing.} \label{Monopolez05z2}
\end{center}
\end{figure}

These features, consisting of oscillations and, at the largest scales $k\lesssim 10^{-3}\, h/\mathrm{Mpc}$, a rise in amplitude, are caused by the non-vanishing, non-divergent parts of the relativistic contributions. To better analyze these effects, it is practical to remove the divergent terms, as they cancel in the full expression and have no impact on the total power spectrum. For that purpose, we define the non-divergent, ``IR-safe'' power spectra, 
\beeq
P^{\text{IR-safe}}(k)\equiv P(k)-P^{\,\text{div.}}(k)\,.
\eneq
According to equations~\eqref{Ps}--\eqref{Pnl},~\eqref{pk0s},~\eqref{Podiv} and~\eqref{pk0nl}, this yields 
\begin{align}
& P_s^{\text{IR-safe}}(k;l=0)=P_z(k;l=0)=\rbr{b^2+\frac 23bf+\frac 15 f^2}D^2P_m(k)\,, \qquad P_o^{\text{IR-safe}}(k;l=0)=0\,, \nnn
&P_{nl}^{\text{IR-safe}}(k;l=0)=\int_0^{\bar r_z}\mathrm d\bar r_{1}\int_0^{\bar r_z}\mathrm d\bar r_{2} \,  \bigg[     \mathcal E_1 \mathcal E_2 \frac{1}{k^4}\rbr{j_{0}(\Delta x)-1+\frac{1}{6}\Delta x^2} -   \mathcal F_1\mathcal F_2\frac{1}{k^2}\rbr{j_{0}''(\Delta x)+\frac 13} \nnn
&\qquad-2\mathcal E_1\mathcal F_2\frac{1}{k^3}\rbr{j'_{0}(\Delta x)+\frac{1}{3}\Delta x}+ 2 \mathcal E_1\mathcal G_2 \frac{ 1}{k^2}\rbr{ j_{0}(\Delta x) + j_{0}''(\Delta x)-\frac 23} \nnn
&\qquad -2\mathcal G_1\mathcal F_2\frac{1}{k}\rbr{j'_{0}(\Delta x)+j'''_{0}(\Delta x)} +  \mathcal G_1\mathcal G_2  \rbr{ j_{0}(\Delta x)  + 2 j_{0}''(\Delta x) + j_{0}''''(\Delta x)} \bigg] P_m(k)\,.
\end{align}
for the IR-safe contributions to the monopole of the observer, source and non-local power spectrum. For the cross power spectra, the respective IR-safe contributions to the monopole are obtained from the terms proportional to $L_0$ in equations~\eqref{PRso}--\eqref{PRonl}. 
Note that, despite the full cancellation $P_o^{\text{IR-safe}}=0$, the contributions of the IR-safe cross power spectra $P_{s\text{-}o}^{\text{IR-safe}}$ and $P_{o\text{-}nl}^{\text{IR-safe}}$  are non-zero. Hence, the observer terms $\delta_o$ cannot be ignored. Furthermore, note that the relativistic contributions that affect the total power spectrum $P_g$ are not individually measurable. Hence, as the divergent contributions cancel out in $P_g$, there is no loss of any physically meaningful quantity when we remove the divergent parts from the individual contributions, while it makes the relativistic features easier to interpret. In figure~\ref{Monopolez05z2}, we have plotted the IR-safe contributions and their sum, the total monopole power spectrum $P_g(k;l=0)$ unaffected by the divergent terms, for redshift $z=0.5$ and, to illustrate the redshift dependence of the relativistic effects, for redshift $z=2$. In these plots, we can easily see that, for redshift $z=0.5$, the elevation of the total power spectrum at the largest scales is caused by the contribution of $P_{s\text{-}o}$, while for $z=2$, the contribution of $P_{s\text{-}nl}$ and $P_{o\text{-}nl}$ are also relevant at these scales. 

\begin{figure}
\begin{center}
\includegraphics[width=7.5cm]{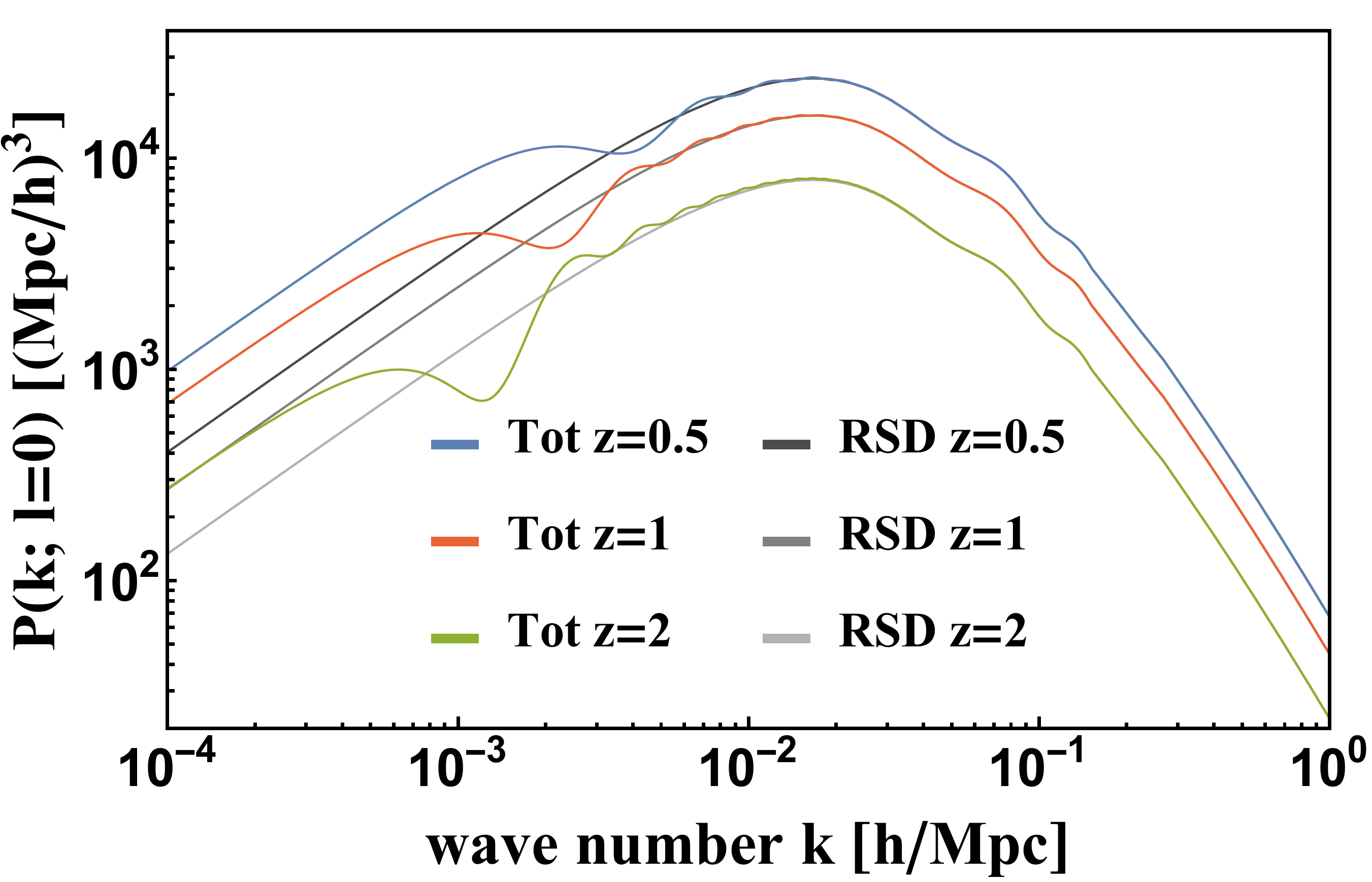}\hspace{0.5cm}  \includegraphics[width=7.5cm]{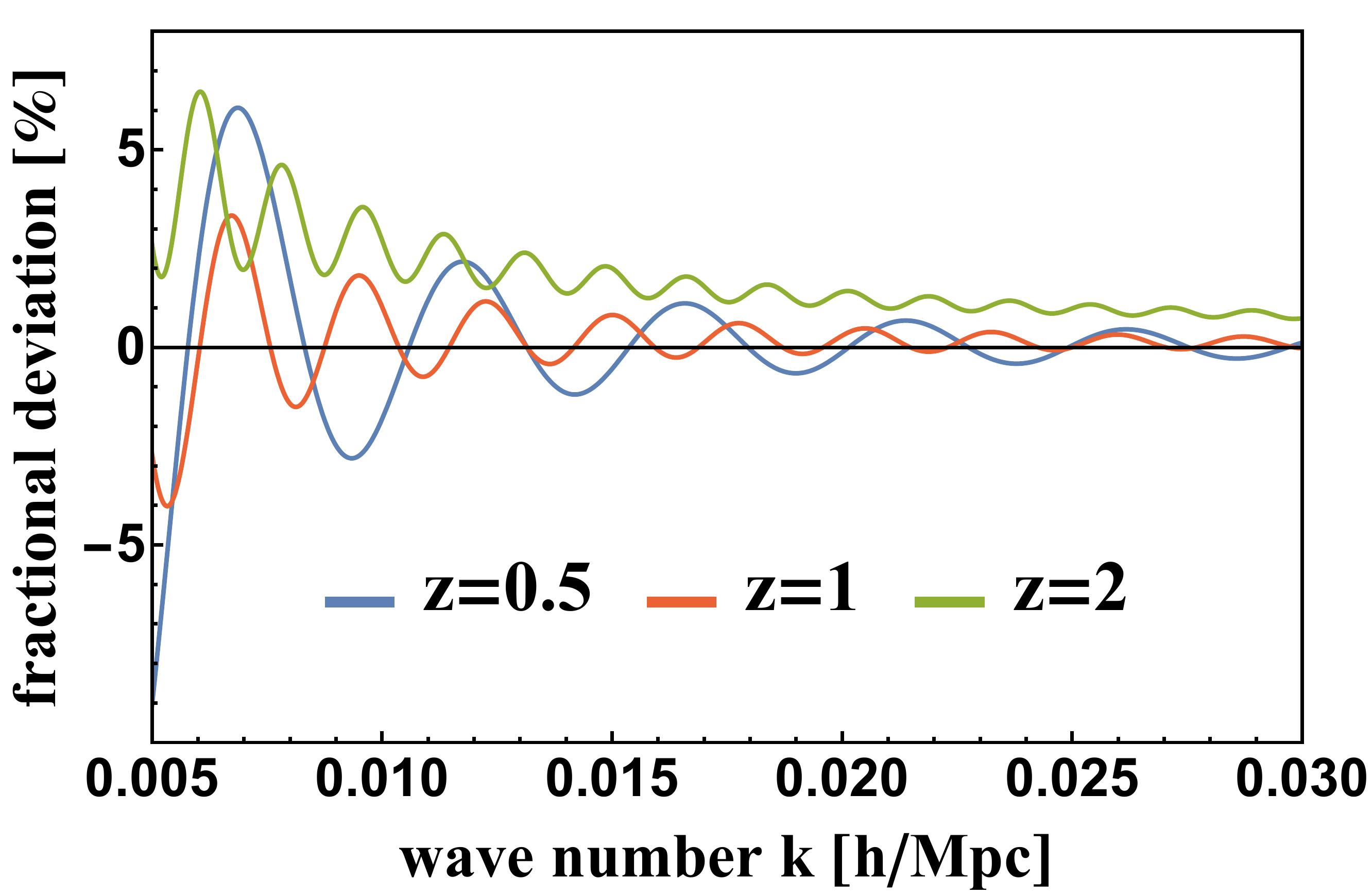}
\caption{On the left, we show the standard redshift-space prediction for the monopole power spectrum $P_z(k;l=0)$ (gray curves) and the total $P_g(k;l=0)$ (colored curves) including all general relativistic effects for $z\in\{0.5,1,2\}$. On the right, we plot the fractional deviation of $P_g(k;l=0)$ from $P_z(k;l=0)$ for  $k=0.005-0.030\,h/\text{Mpc}$.}  \label{P0fracdev}
\end{center}
\end{figure}

At scales $k\gtrsim 10^{-3}\,h/\mathrm{Mpc}$, the source-observer cross power spectrum $P_{s\text{-}o}$, which is the most significant contribution at low redshift $z=0.5$, shows an oscillating behavior due to the appearance of spherical Bessel function which are not integrated over. These oscillations are visible in the total monopole power spectrum at scales $k\lesssim k_{eq}$, before the relativistic effects become negligible compared to the standard redshift-space prediction. For $z=2$, the width of these oscillations is smaller (despite their appearance in the logarithmic plot) and they occur at lower $k$ due to the higher value of $\rz$ in the argument of the spherical Bessel function $j_l(k\rz)$ and its derivatives. The heights of the positive and negative peaks are determined by the behavior of the functions $\mathcal A$, $\mathcal B$, $\mathcal C$, $\mathcal D$ and $D$ shown in figure~\ref{z-functions}. At redshift $z=0.5$, the dominant contribution to $P_{s\text{-}o}$ arises from the last two terms in equation~\eqref{Pso}, those proportional to $D\mathcal D$. The other terms are less significant, and in particular those proportional to $\mathcal A$ are completely negligible. However, this changes for the redshift $z=2$, where these terms $\propto\mathcal A$  drastically increase in significance. In particular, the term $\propto\mathcal A\mathcal D$ now yields a negative contribution which significantly decreases the amplitude of the positive peaks while increasing the one of the negative peaks, as visible in the lower panel of figure~\ref{Monopolez05z2}.

While at redshift $z=0.5$, the contribution of $P_{s\text{-}o}$ is the most significant alteration to the standard redshift-space monopole power spectrum, the contributions of $P_{nl}$, $P_{s\text{-}nl}$ and $P_{o\text{-}nl}$, which contain integrations over the line-of-sight, gain importance with higher redshift $z$. These contributions, in particular $P_{o\text{-}nl}$, and $P_{nl}$ at $k\gtrsim 5\times 10^{-3}$, result in an increase of the total power spectrum. This behavior is well visible in figure~\ref{P0fracdev}, in particular in the right panel. At redshift $z=0.5$, where the contributions involving non-local terms have little significance, $P_{s\text{-}o}$ causes oscillations around the reference level of the standard redshift-space power spectrum, with similarly prominent negative and positive peaks. At redshift $z=2$, where the contribution of $P_{nl}$ even exceeds the one of $P_{s\text{-}o}$ at $k\gtrsim 8\times 10^{-3}$, these oscillations are clearly lifted above the reference level of $P_z$. For further comparison, figure~\ref{P0fracdev} also contains an intermediate redshift $z=1$, where the elevation caused by the non-local terms is clearly visible, while not being as prominent as for redshift $z=2$. Note that, at $k\approx 5\times 10^{-3}h/\mathrm{Mpc}$, the redshift $z=0.5$ shows the largest deviation from the standard redshift-space prediction, since the oscillations occur at higher $k$ for lower $z$, as visible on the left panel of the figure. However, for $k\approx 3\times 10^{-2}h/\mathrm{Mpc}$, where the oscillations induced by $P_{s\text{-}o}$ have little significance, the largest deviation occurs at the highest redshift $z=2$ due to the impact of $P_{nl}$.

From the right panel of figure~\ref{P0fracdev}, we further conclude that the relativistic corrections are still at the percent-level at $k\approx 0.02\,h/\text{Mpc}$, roughly corresponding to the scale of matter-radiation equality. Thus, while relativistic effects are negligible at small scales, with $k$ considerably larger than $k_{eq}$, they must be taken into account for analyzing data on large scale, $k\lesssim k_{eq}$. 

\subsection{Quadrupole} \label{quadrupole}
 \begin{figure}
\begin{center}\includegraphics[width=12cm]{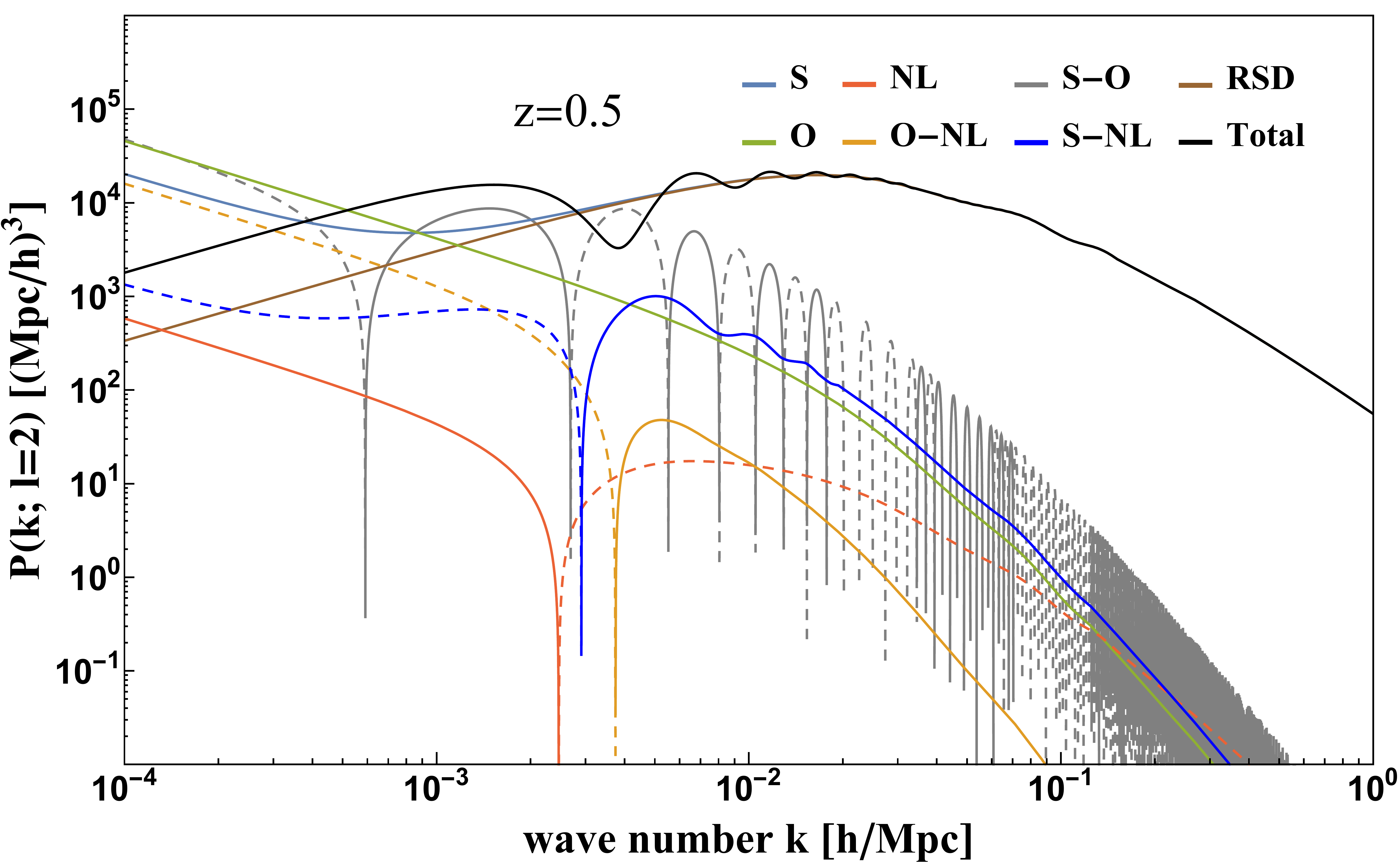}\end{center}
\caption{Same as figure~\ref{P0z05div}, but for the quadrupole $l=2$.} \label{P2z05div}
\end{figure}
Similarly as for the monopole, we numerically evaluate the contributions to the quadrupole power spectrum, given by the terms proportional to $L_2$ in equations~\eqref{Ps}--\eqref{Pnl} and~\eqref{Pso}--\eqref{Ponl}, and show the resulting curves for redshift $z=0.5$ in figure~\ref{P2z05div}. Again, we see that the individual contributions to the quadrupole, in particular the contribution of the source power spectrum $P_s$, exhibit a divergent behavior in the infrared, which vanishes for the total quadrupole power spectrum $P_g(k; l=2)$ as discussed in section~\ref{IRdiv}. Comparing $P_g(k; l=2)$ (black curve) with the standard redshift-space quadrupole power spectrum (brown curve), we see prominent additional features at scales $k\lesssim k_{eq}$, consisting out of oscillations and a rise in amplitude at the largest scales $k\lesssim 10^{-3}h/\mathrm{Mpc}$. 

\begin{figure}\begin{center}
\includegraphics[width=12cm]{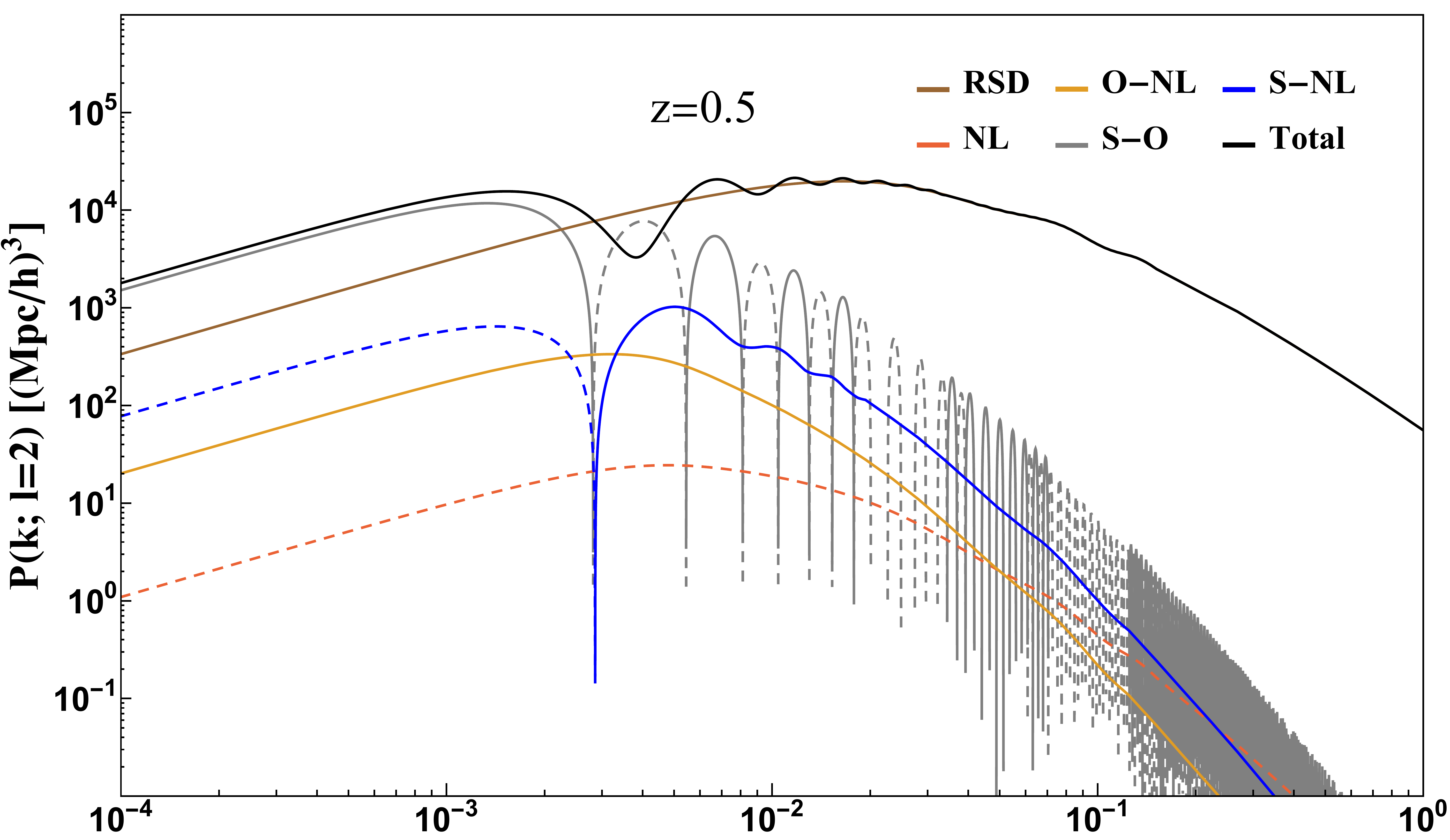} 
\includegraphics[width=12cm]{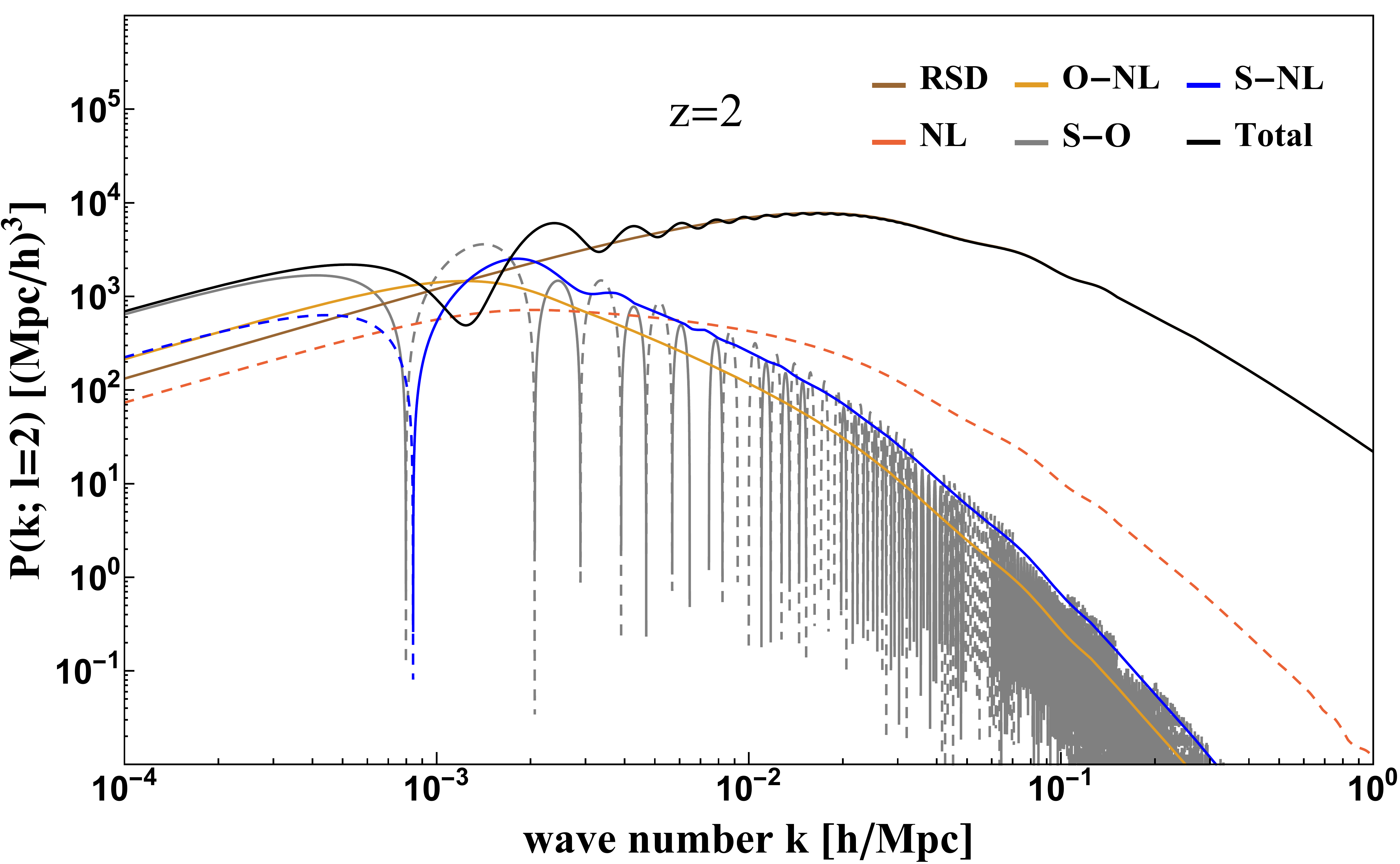}
\caption{Same as figure~\ref{Monopolez05z2}, but for the quadrupole $l=2$.} \label{Quadrupolez05z2}
\end{center}
\end{figure}

Again, to better analyze these features, we consider the IR-safe contributions, given by
\begin{align}
& P_s^{\text{IR-safe}}(k;l=2)=P_z(k;l=2)=\rbr{\frac 43 bf+\frac 47f^2}D^2P_m(k)\,,\qquad P_o^{\text{IR-safe}}(k;l=2)=0\,, \nnn
&P_{nl}^{\text{IR-safe}}(k;l=2)=-5\int_0^{\bar r_z}\mathrm d\bar r_{1}\int_0^{\bar r_z}\mathrm d\bar r_{2} \,  \bigg[     \mathcal E_1 \mathcal E_2 \frac{1}{k^4}\rbr{j_{2}(\Delta x)-\frac{1}{15}\Delta x^2} -   \mathcal F_1\mathcal F_2\frac{1}{k^2}\rbr{j_{2}''(\Delta x)-\frac{2}{15}} \nnn
&\qquad-2\mathcal E_1\mathcal F_2\frac{1}{k^3}\rbr{j'_{0}(\Delta x)-\frac{2}{15}\Delta x}+ 2 \mathcal E_1\mathcal G_2 \frac{ 1}{k^2}\rbr{ j_{2}(\Delta x) + j_{2}''(\Delta x)-\frac{2}{15}} \nnn
&\qquad -2\mathcal G_1\mathcal F_2\frac{1}{k}\rbr{j'_{2}(\Delta x)+j'''_{2}(\Delta x)} +  \mathcal G_1\mathcal G_2  \rbr{ j_{2}(\Delta x)  + 2 j_{2}''(\Delta x) + j_{2}''''(\Delta x)} \bigg] P_m(k)\,, \label{IRsafepowerspectra}
\end{align}
and the terms proportional to $L_2$ in the equations~\eqref{PRso}--\eqref{PRonl} for the cross-power spectra. The numerical results for the IR-safe power spectra are shown in figure~\ref{Quadrupolez05z2} for the redshift $z=0.5$ (top panel) and $z=2$ (bottom panel). For the redshift $z=0.5$, we see that the oscillating contribution of $P_{s\text{-}o}$ is the only significant alteration to the standard prediction. At the higher redshift $z=2$, we see that all the other contributions, which involve line-of-sight integrals, have gained significance and lead to additional alterations apart from the oscillations induced by $P_{s\text{-}o}$. In particular, at $k\lesssim 7\times 10^{-3} h/\mathrm{Mpc}$, the contribution of $P_{s\text{-}nl}$ leads to an increase in the total quadrupole power spectrum, while the contribution of $P_{nl}$ leads to a decrease at $k\gtrsim 7\times 10^{-3} h/\mathrm{Mpc}$.  

\begin{figure}
\begin{center}
\includegraphics[width=7.5cm]{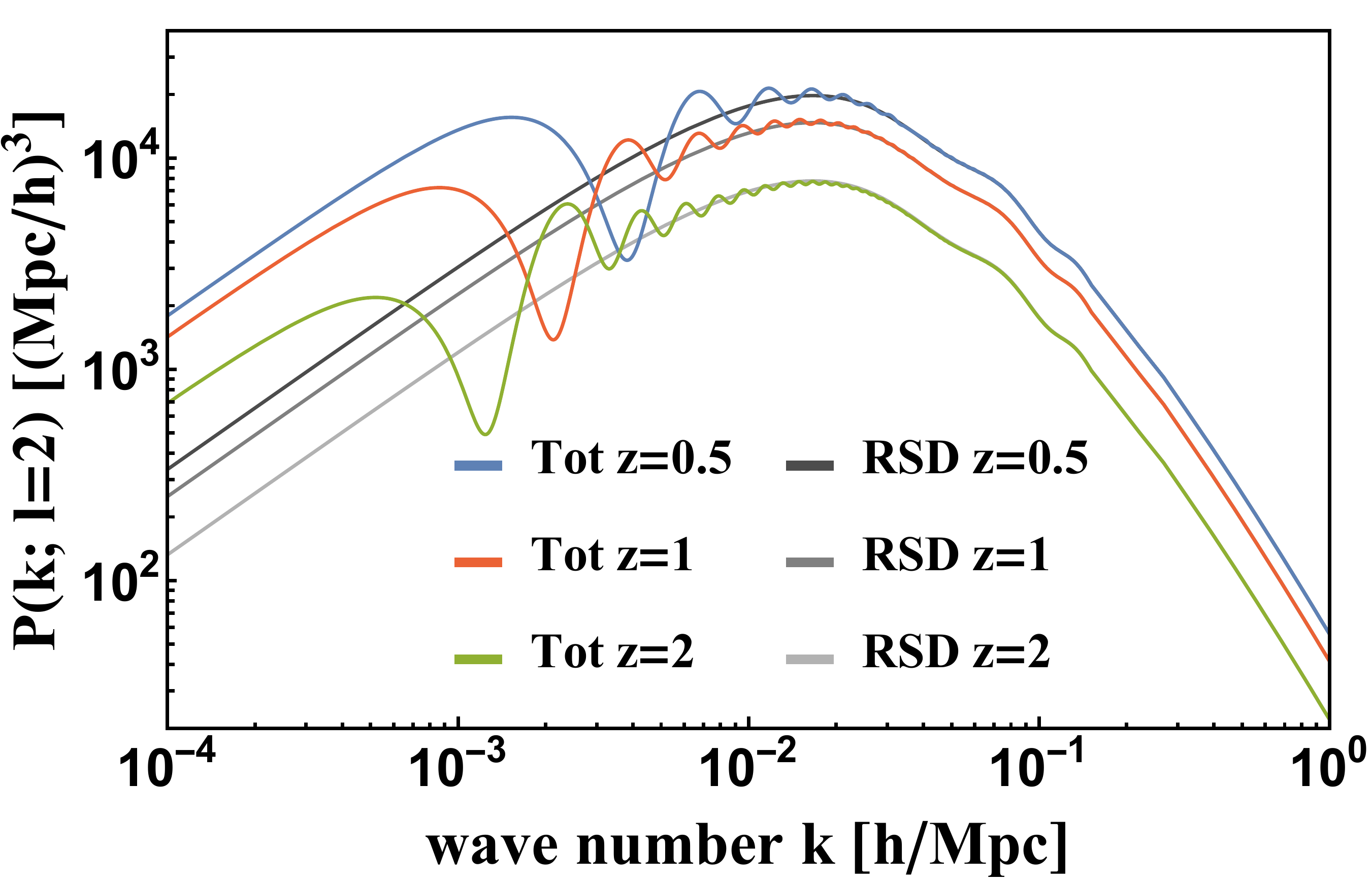}\hspace{0.5cm}  \includegraphics[width=7.5cm]{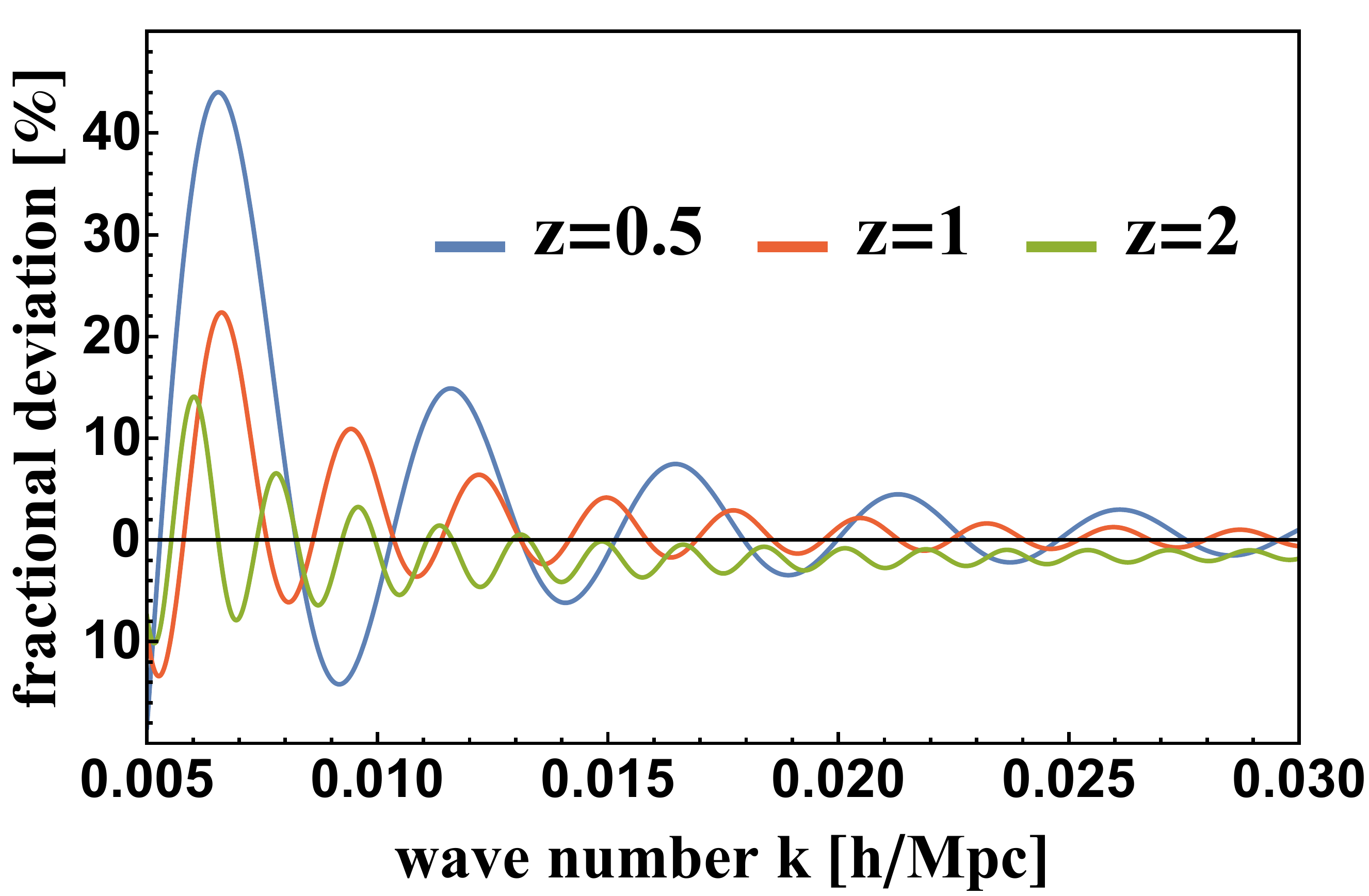}
\caption{Same as figure~\ref{P0fracdev}, but for the quadrupole $l=2$.}  \label{P2fracdev}
\end{center}
\end{figure}

Compared to the monopole power spectrum, the oscillations induced by $P_{s\text{-}o}$ have an even higher effect onto the total quadrupole higher spectrum, which is due to the overall factor of $5$ in front of the terms proportional to $L_2$ in equation~\eqref{PRso}, arising from the factor $(2l+1)$ in the plane wave expansion~\eqref{planewave}. This is clearly visible on the right panel of figure~\ref{P2fracdev}, showing a deviation at $k\approx 5\times 10^{-3}h/\mathrm{Mpc}$ of more than $10\%$ for $z=2$ (more than $20\%$ for $z=1$, and more than $40\%$ for $z=0.5$). Comparing this with the right panel of figure~\ref{P0fracdev}, the deviation of the total monopole power spectrum from the standard redshift-space prediction is below $10\%$ for all three redshifts. At $k\approx 3\times 10^{-2} h/\mathrm{Mpc}$, the deviation is, with about $2\%$, the largest for the redshift $z=2$ due the contribution of $P_{s\text{-}nl}$.

With even higher deviations from the standard redshift-space prediction than for the monopole, this analysis of the impact of general relativistic effects on the quadrupole power spectrum reaffirms our conclusions that these effects need to be taken into account at scales $k\lesssim k_{eq}$.

\subsection{Hexadecapole} \label{hexadecapole}

As discussed in section~\ref{IRdiv}, IR-divergent terms only appear in the monopole and quadrupole power spectra, meaning that all individual contributions to the hexadecapole are already IR-safe. In particular, according to equations~\eqref{Ps}--\eqref{Po}, the source power spectrum is equal to the standard redshift-space prediction, and the observer power spectrum is fully vanishing,
\beeq
P_s(k; l=4)=P_z(k;l=4)=\frac{8}{35}f^2D^2P_m(k)\,,\qquad P_o(k;l=4)=0\,.
\eneq 
\begin{figure}\begin{center}
\includegraphics[width=12cm]{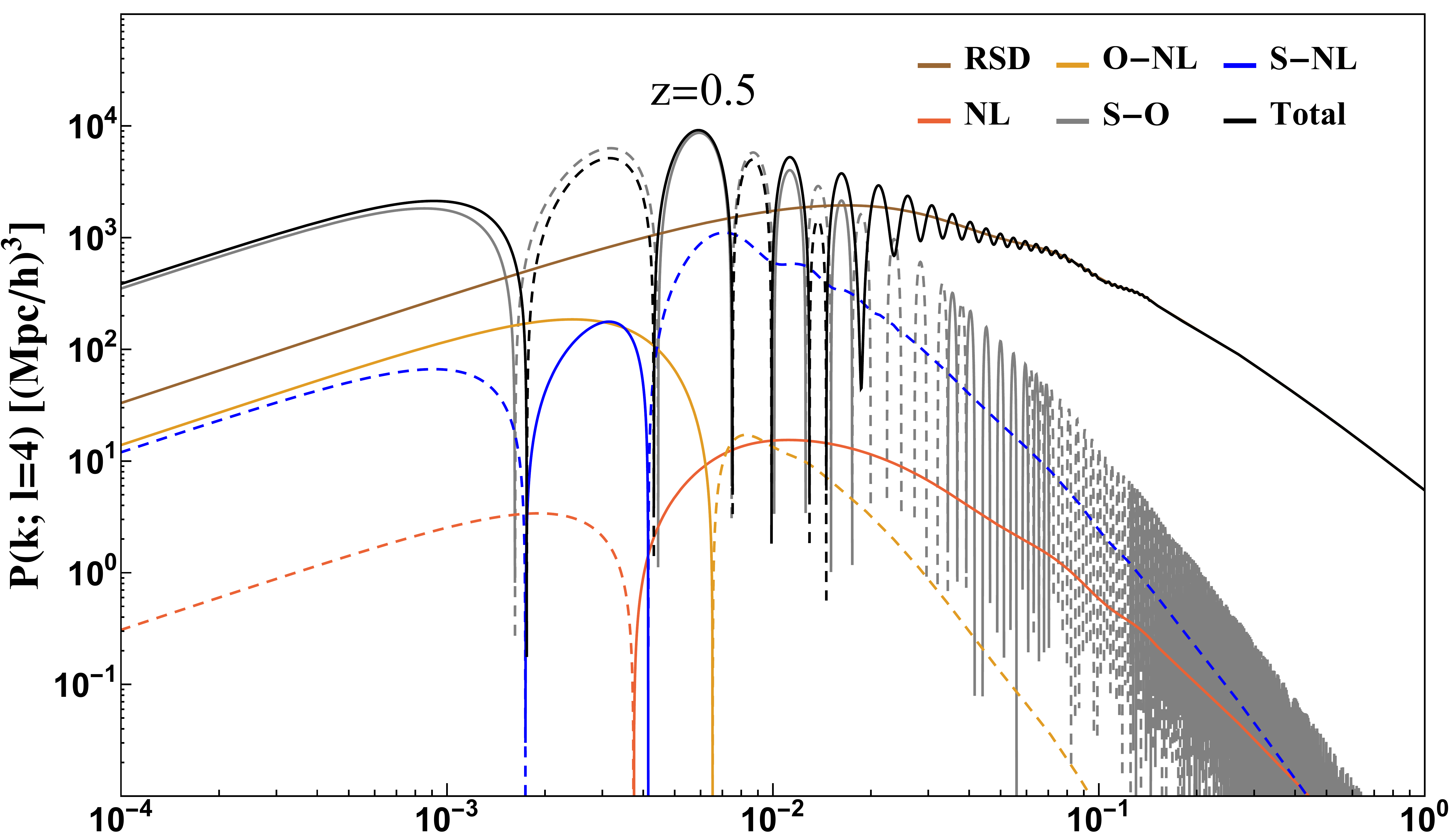} 
\includegraphics[width=12cm]{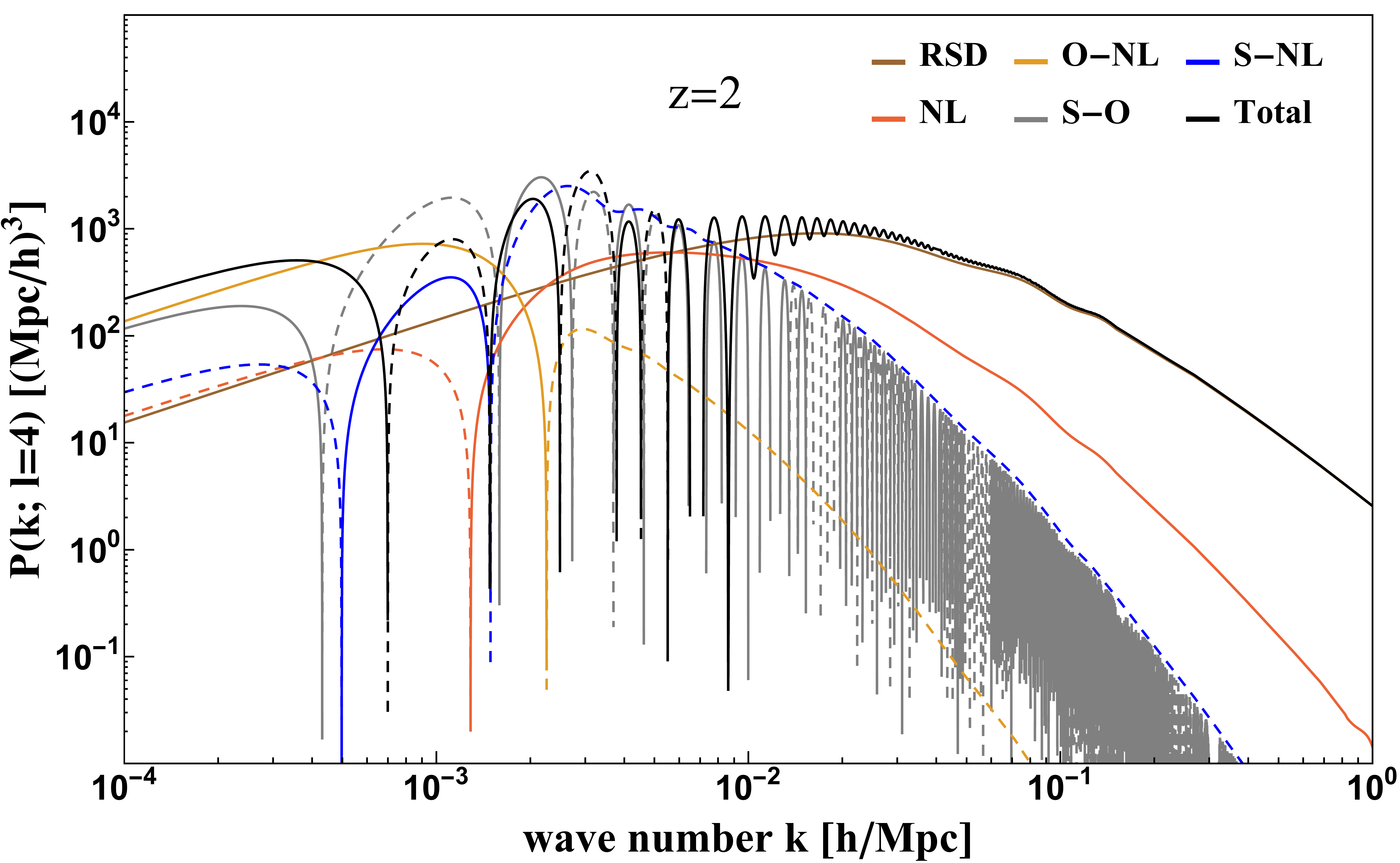}
\caption{This plot show the total hexadecapole power spectrum and its individual components at redshifts $z=0.5$ (top) and $z=2$ (bottom). Note that none of the individual contributions exhibit an IR-divergent behavior. In particular, the $P_o(k;l=4)$ is exactly vanishing and $P_s(k;l=4)$ is identical to the standard redshift-space prediction, which is why there are no curves for these contributions.} \label{Hexadecapolez05z2}
\end{center} 
\end{figure} 

\begin{figure}
\begin{center}
\includegraphics[width=10cm]{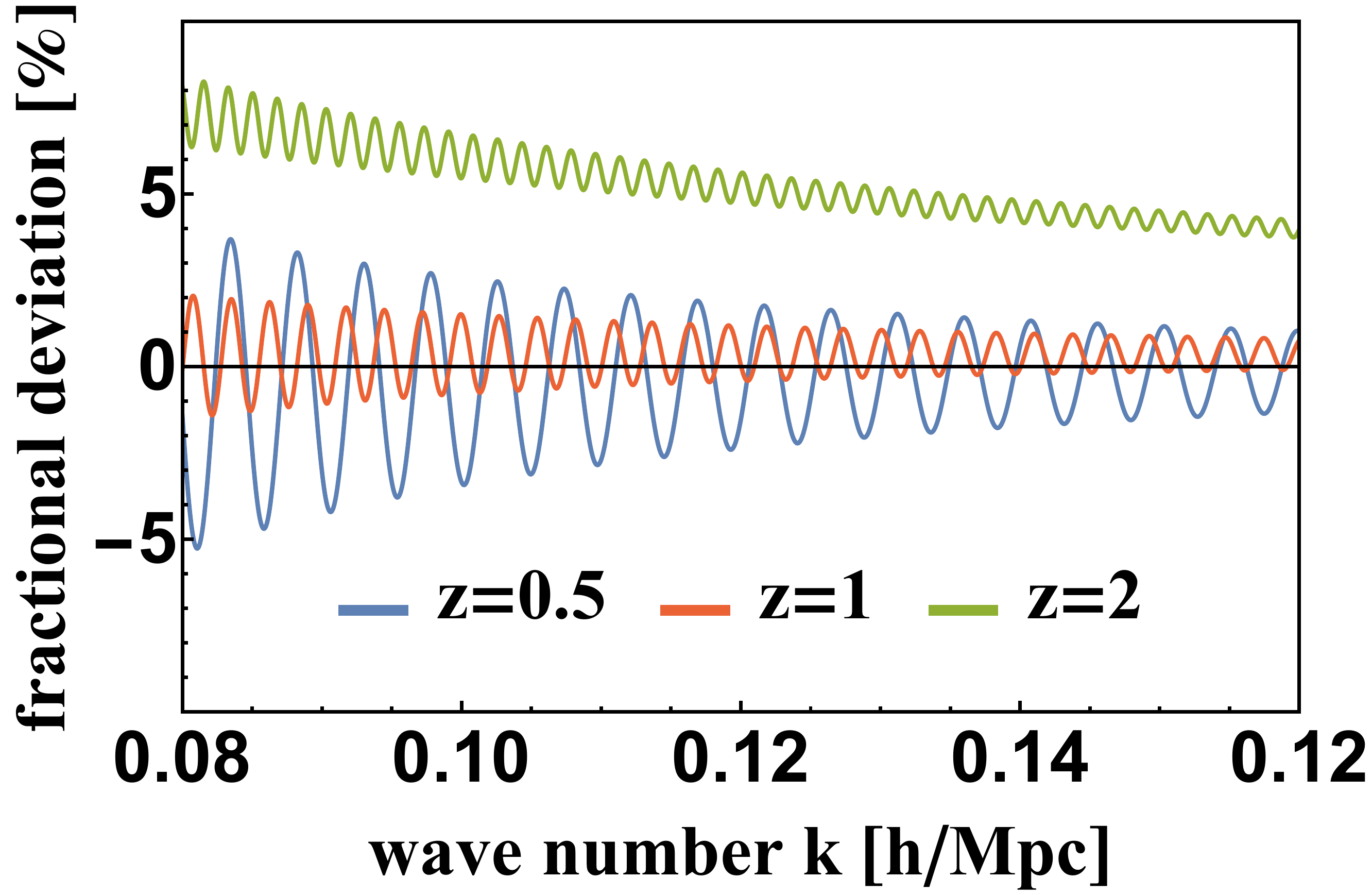}
\caption{We plot the fractional deviation of $P_g(k;l=4)$ from $P_z(k;l=4)$ for  $k=0.08-0.12\,h/\text{Mpc}$. Note that, compared to the monopole and quadrupole, the relativistic corrections have an even higher significance and alter the total hexadecapole power spectrum up to higher values of $k$, which is why we chose a higher range of $k$ compared to figures~\ref{P0fracdev} and~\ref{P2fracdev}.}  \label{P4fracdev}
\end{center}
\end{figure}

\noindent The contributions of the source power spectrum and the cross power spectra can be obtained by taking the terms proportional to $L_4$ in equations~\eqref{Pnl} and~\eqref{Pso}--\eqref{Ponl}. Following these equations, we evaluate the quadrupole power spectrum with its individual contributions numerically for the redshifts $z=0.5$ and $z=2$ and show the results in figure~\ref{Hexadecapolez05z2}. As for the monopole and quadrupole, we see that at the lower redshift $z=0.5$, the clearly most significant relativistic correction to the standard redshift-space prediction arises from the source-observer cross-power spectrum $P_{s\text{-}o}$. However, as the standard redshift-space power spectrum is about an order of magnitude lower for the hexadecapole $l=4$ than for the monopole $l=0$ or quadrupole $l=2$, the oscillations induced by $P_{s\text{-}o}$ have a drastically larger impact on the total hexadecapole power spectrum. At $k\lesssim 10^{-2}h/\mathrm{Mpc}$, the total power spectrum is dominated by the contribution of $P_{s\text{-}o}$, and the oscillations are still clearly visible in the total curve up to $k\approx 10^{-1}h/\mathrm{Mpc}$. Indeed, as seen in figure~\ref{P4fracdev}, the deviation from the standard redshift-space prediction is still a few percent at that scale. 

For the higher redshift $z=2$, we see that all other relativistic corrections involving integrals over the line-of-sight have considerably gained significance, and affect the total hexadecapole power spectrum on some scales: At scales $k\gtrsim 10^{-2}h/\mathrm{Mpc}$, the positive contribution of $P_{nl}$ leads to an increase in the total curve, while at larger scales $10^{-3}h/\mathrm{Mpc}\lesssim k \lesssim 10^{-2}h/\mathrm{Mpc}$, the negative contribution of $P_{s\text{-}nl}$ decreases the positive while increasing the negative peaks. Indeed, the elevating impact of $P_{nl}$ is clearly visible in figure~\ref{P4fracdev}, with the deviation from the standard redshift-space prediction still being around $5\%$ at $k=1.2 \times 10^{-1}h/\mathrm{Mpc}$.  At the largest scales $k\lesssim 10^{-3}h/\mathrm{Mpc}$, the contribution of $P_{o\text{-}nl}$ yields an additional rise in amplitude. 

These results illustrate that when considering the hexadecapole, taking into account all relativistic effects in absolutely crucial, even at rather small scales $k\approx 10^{-1}h/\mathrm{Mpc}$. In principle, this is also true for higher multipoles, $l=6,8,\dots$, as the standard redshift-space power spectrum is completely vanishing for $l\geq 6$, while $P_{s\text{-}o}$, $P_{nl}$, $P_{s\text{-}nl}$ and $P_{o\text{-}nl}$ have contributions to all these higher even multipoles. However, these contributions decline in significance with increasing $\ell$ due to the behavior of the spherical Bessel functions $j_\ell(x)$, with negligible values at large $x$ and the values at small $x$ decreasing with increasing $\ell$.

\section{Summary and conclusion}\label{conclusion}

In this paper, we have first discussed the relation between the observed and the theory galaxy power spectrum, and then studied the general relativistic contributions 
in galaxy clustering based on the latter.
The observed power spectrum is obtained by Fourier transforming the observed galaxy fluctuation, with the (finite) observed volume located on the past light cone as an integration domain. In contrast, the theory power spectrum is defined on an infinite hypersurface of simultaneity set by a single conformal time coordinate within the redshift range of the observed survey volume. Since we have access only to the past light-cone volume, the theory power spectrum 
itself is not directly measurable. However, its convolution with the survey geometry over the redshift range, as described by equation~\eqref{power_spectrum_obs_k'=k}, is related to the observed power 
spectrum. More importantly, as it is defined on
a hypersurface, it is independent of individual survey geometry. Therefore, it is highly suitable to study the impact of general relativistic effects on a theoretical level. 
In particular, the theory power spectrum naturally incorporates the relativistic contributions along
the line-of-sight direction such as the gravitational lensing effect, which
is hard to account for in the standard power spectrum analysis. When only the redshift-space distortion is considered, it reduces to the well-known standard redshift-space power spectrum. 

We use the theory power spectrum to analyze the full
relativistic contributions in the observed galaxy fluctuation from
the source position and along the line-of-sight direction, including 
the observer position.
In the past, there have been considerable efforts to derive the galaxy power spectrum accounting for relativistic effects (see, e.g., \cite{Jeong,Yoo:2010,Yoo:2012}). 
However, these studies considered only the contribution at the source position
$\delta_s$, ignoring the
contributions $\delta_o$ at the observer position and~$\delta_{nl}$ along
the line-of-sight direction. While $\delta_s$ is the dominant contribution on small scales, the other contributions become comparable on large scales.
Consequently, ignoring $\delta_o$ and $\delta_{nl}$ results in significant
systematic errors in the power spectrum computation on large scales.

In particular, the galaxy power spectrum diverges at large scales when the relativistic contribution from only $\delta_s$ is considered.
This infrared divergence is considered in literature as a contaminant
to the measurements of the primordial non-Gaussianity (e.g., \cite{Jeong, nG1, nG2, nG3, nG4, nG5, nG6}).
However, it is an {\it artifact} 
of inadvertently ignoring the other relativistic
contributions~$\delta_o$ and~$\delta_{nl}$ and hence breaking the gauge
invariance. Our calculations show that $\delta_o$ and $\delta_{nl}$ also
diverge in the infrared, and indeed all these divergent terms at low~$k$ cancel out, yielding no terms $\propto k^{-4}P_m(k)$ or $\propto k^{-2}P_m(k)$ in the total power spectrum. This statement is valid for all surveys, independent of any specific survey geometry.
As we have demonstrated, the vanishing of the infrared divergence is in accordance with the equivalence principle in general 
relativity: A perturbation acting as a uniform
gravitational force on the scale of the survey cannot have any impact on the measured quantities. Though not investigated in this work, we suspect that a similar argument can be made for the local-type non-Gaussian signature in the initial condition, and hence for its impact on the galaxy power spectrum initially discovered in~\cite{nG7}. 

While no diverging power is present on large scales, relativistic corrections represented by non-diverging terms alter the galaxy power spectrum. As the standard matter power spectrum falls over beyond $k_{eq}$, these corrections cause the full relativistic galaxy power spectrum to significantly differ from its standard redshift-space prediction at scales $k\lesssim k_{eq}$. Our numerical investigation of the total monopole, quadrupole and hexadecapole power spectra show that, at the largest scales $k\lesssim 10^{-3} h/\mathrm{Mpc}$, the relativistic corrections lead to a rise in amplitude. At scales $10^{-3} h/\mathrm{Mpc}\lesssim k\lesssim 10^{-2} h/\mathrm{Mpc}$, they induce oscillating features arising from the cross power spectrum $P_{s\text{-}o}$ of the observer and source terms. In addition, the contribution of non-local terms integrated along the line-of-sight become relevant at high redshift, further altering the amplitude of the total power spectrum. These effects result, for the monopole and quadrupole, in a percent-level deviation of the relativistic power spectrum from its standard redshift-space prediction at scales $k\approx k_{eq}$. When additionally considering the hexadecapole, where the amplitude of the standard redshift-space power spectrum is lower compared to the monopole and quadrupole, taking relativistic corrections into account is even more important as they significantly alter the total hexadecapole power spectrum up to rather small scales $k\approx 10^{-1}h/\mathrm{Mpc}$.

We conclude that, while relativistic effects are negligible on small scales $k\gg k_{eq}$, taking them into account is crucial for any survey targeting large scales, $k\lesssim k_{eq}$. While the relativistic theory power spectrum analyzed in this work is independent of any particular survey, it is related to the observed power spectrum with a given survey geometry via equation~\eqref{power_spectrum_obs_k'=k}. Hence, our work does not only point out the significance of relativistic effects on a theoretical level, but also provides all fundamental tools to determine the relativistic galaxy power spectrum for any specific survey, which will be investigated in future work.

\acknowledgments

We thank Ruth Durrer, Ermis Mitsou and Enea Di Dio for useful discussions. We acknowledge support by the Swiss National Science Foundation (SNF CRSII5\_173716), and J.Y.~is further supported by a Consolidator Grant of the European Research Council (ERC-2015-CoG grant 680886).

\appendix

\section{Cross power spectra of the local and non-local contributions to the observed galaxy fluctuation}\label{A}

Here, we provide the expressions for the cross power spectra of the contributions $\delta_s$, $\delta_o$, $\delta_{nl}$ to the observed galaxy fluctuation $\delta_g$ in equations~\eqref{dg}$-$\eqref{nl}. We present the full expressions for the cross power spectra $P_{s\text{-}o}$, $P_{s\text{-}nl}$ and $P_{o\text{-}nl}$, expanding them into angular multipoles with respect to the cosine angle $\mu_k=\boldsymbol{\hat k}\cdot\boldsymbol{\hat n}$. 
Since the power spectra are real and symmetric under rotation around $\boldsymbol{\hat n}$, the expansion contains only even multipoles.
As discussed in section~\ref{IRdiv}, the monopole and quadrupole contributions to the individual power spectra diverge in the infrared. However, the divergent parts cancel out when all individual monopole and quadrupole power spectra are summed. Therefore, after isolating the divergent parts, we present the expressions for the IR-safe cross power spectra.

The cross power spectra of the contributions $\delta_s$, $\delta_o$, $\delta_{nl}$ to the density fluctuation $\delta_g$ are obtained from the variances  $\langle \delta_s^{}  \delta_o^* \rangle$, $\langle \delta_s^{}  \delta_{nl}^* \rangle$, $\langle \delta_o^{}  \delta_{nl}^* \rangle$ by using the method described in section~\ref{contribution to PS}. Note that, as stated in equation~\eqref{Ptot}, these cross power spectra contribute to the total power spectrum $P_g(\bm k)$ with a factor of 2, which we account for in all figures of section~\ref{numerical}.  
We obtain
\begin{align}
\frac{ P_{s\text{-}o}(\bm k)}{P_m(k)} = \sum_{n=0}^\infty &(-1)^n (4n+1)L_{2n} 
  \bigg[  \mathcal A \mathcal C \frac{1}{k^4}  j_{2n}(k\bar r_z)
	 + \rbr{ \mathcal B \mathcal C   -\mathcal A\mathcal D }   \frac{1}{k^3}  j_{2n}'(k\bar r_z) 
		 + b D \mathcal C  \frac{1}{k^2}  j_{2n}(k\bar r_z) \nnn
&- \rbr{ \mathcal B \mathcal D +    fD\mathcal C }   \frac{1}{k^2}  j_{2n}''(k\bar r_z) 
	{-} b D \mathcal D  \frac{1}{k}  j_{2n}'(k\bar r_z)
		 + f D  \mathcal D  \frac{1}{k} j_{2n}'''(k\bar r_z)\bigg] \,,\label{Pso}
\end{align}
for the source-observer cross power spectrum,
\begin{align}
\frac{P_{s\text{-}nl}(\bm k)}{P_m(k) }  =  \sum_{n=0}^\infty& (-1)^n (4n+1)L_{2n}  
  \int_0^{\bar r_z}\mathrm d\bar r\, \bigg[ \mathcal A  \mathcal E   \frac{1}{k^4}j_{2n}(\Delta x_z) 
 +\rbr{\mathcal B \mathcal E   -\mathcal A \mathcal F }   \frac{ 1}{k^3}j_{2n}'(\Delta x_z)    \nnn
&+b D  \mathcal E   \frac{ 1}{k^2}j_{2n}(\Delta x_z)  + \mathcal A  \mathcal G   \frac{ 1 }{k^2}\rbr{j_{2n}(\Delta x_z) +  j_{2n}''(\Delta x_z)}   
-\rbr{ D f \mathcal E  + \mathcal B\mathcal F } \frac{1}{k^2}j_{2n}''(\Delta x_z)  \nnn
&+ D f \mathcal F  \frac{ 1}{k} j_{2n}'''(\Delta x_z) - b D \mathcal F  \frac{1}{k} j_{2n}'(\Delta x_z)   +  \mathcal B \mathcal G   \frac{1 }{k}\rbr{ j_{2n}'(\Delta x_z) +  j_{2n}'''(\Delta x_z)} \nnn
&  + b D   \mathcal G \,  \rbr{  j_{2n}(\Delta x_z) +  j_{2n}''(\Delta x_z)}- D f \mathcal G\rbr{ j_{2n}''(\Delta x_z) + j_{2n}''''(\Delta x_z)} \bigg]\label{Psnl}
  \,,
\end{align}
for the source-non local cross power spectrum, and
\begin{align}
\frac{P_{o\text{-}nl}(\bm k) }{ P_m(k)}  =   \sum_{n=0}^\infty &(-1)^n (4n+1)L_{2n}  
 \int_0^{\bar r_z}\mathrm d\bar r\, \bigg[    \mathcal C\mathcal E    \frac{1}{k^4}  j_{2n}(k\bar r)
		   -  \rbr{ \mathcal D \mathcal E   - \mathcal C\mathcal F}  \frac{1}{k^3}  j_{2n}'(k\bar r)- \mathcal D \mathcal F   \frac{1}{k^2}  j_{2n}''(k\bar r)  \nnn
& +  \mathcal C	 \mathcal G     \frac{1}{k^2} \rbr{  j_{2n}(k\bar r) +  j_{2n}''(k\bar r)}
- \mathcal D \mathcal G \frac{1}{k}  \rbr{ j_{2n}'(k\bar r)  + j_{2n}'''(k\bar r)}\bigg] \,,\label{Ponl}
\end{align}
for the observer-non local cross power spectrum, where the Legendre polynomials are functions of $\mu_k$ and we defined $\Delta r_z\equiv \bar r_z-\bar r$ and $\Delta x_z\equiv k \Delta r_z$. We now isolate the infrared-divergent terms in the above cross power spectra, which yields
\begin{align}
\frac{P_{s\text{-}o}^{\,\text{div.}}(\bm k) }{P_m(k)}
	  = &  \mathcal A \mathcal C \frac{1}{k^4}  \rbr{ \rbr{ 1  - \frac 16  k^2 \bar r_z^2 }L_0  - \frac 13 k^2 \bar r_z^2  L_2  }+ b D \mathcal C\frac{1}{k^2} L_0  
		 \nnn
		 &-\frac 13 \rbr{ \bar r_z 
		 \mathcal B \mathcal C   -\bar r_z \mathcal A\mathcal D - \mathcal B \mathcal D  -  D f \mathcal C}   \frac{1}{k^2}  \rbr{L_0 +2L_2 } \label{Pk0so}\,,
\end{align}
for the IR-divergent part of the observer-source cross power spectrum,
\begin{align}
\frac{P_{s\text{-}nl}^{\,\text{div.}}( \bm k)}{P_m(k)} 	=  \int_0^{\bar r_z}\mathrm d\bar r\,  
		& \bigg[
		 \mathcal A  \mathcal E   \frac{1}{k^4}\rbr{ \rbr{ 1  - \frac 16  \Delta x_z^2}L_0  - \frac 13 \Delta x_z^2  L_2}+ \frac23  \mathcal A  \mathcal G   \frac{1}{k^2}\rbr{L_0 - L_2} \nnn
& - \frac 13 \rbr{ \Delta r_z (\mathcal B \mathcal E   -  \mathcal A \mathcal F ) - D f \mathcal E  -  \mathcal B\mathcal F }  \frac{1}{k^2} \rbr{L_0 +2L_2}
  + b D  \mathcal E  \frac{1}{k^2}  L_0 \bigg] \,,
\end{align}
for the IR-divergent part of the observer-non local cross power spectrum, and
\begin{align}
\frac{P_{o\text{-}nl}^{\,\text{div.}}( \bm k)}{P_m(k)} =  \int_0^{\bar r_z}\mathrm d\bar r\,& \bigg[
		 \mathcal C\mathcal E    \frac{1}{k^4}   \rbr{ \rbr{ 1  - \frac 16  k^2 \bar r^2 }L_0  - \frac 13 k^2 \bar r^2  L_2 }+
	 \frac 23 \mathcal C	 \mathcal G   \frac{1}{k^2} \rbr{L_0 -L_2 }
 \nnn
 &
  + \frac 13\rbr{ \bar r  \mathcal D \mathcal E    - \bar r  \mathcal C\mathcal F  +  \mathcal D \mathcal F }  \frac{1}{k^2} \rbr{L_0 +2L_2}     \bigg] 
 \,,\label{Pk0onl}
\end{align}
for the IR-divergent part of the source-non local cross power spectrum. Note that these divergent contributions, given by the terms proportional to $k^{-2}P_m$ and $k^{-4}P_m$, only appear in the monopoles and the quadrupoles of the cross power spectra.

The IR-safe cross power spectra defined as $P^{\text{IR-safe}}(k)\equiv P(k)-P^{\,\text{div.}}(k)$, i.e.~with the divergent terms being removed, are then given by
\begin{align}
&\frac{ P_{s\text{-}o}^{\text{IR-safe}}( \bm k)}{P_m(k)} =   \sum_{n=0}^\infty (-1)^n (4n+1)L_{2n}  \bigg[ \mathcal A \mathcal C \frac{1}{k^4}\rbr{j_{2n}(k\bar r_z)-\rbr{1-\frac{1}{6}k^2\bar r_z^2}\delta_{n0}-\frac{1}{15}k^2\bar r_z^2\delta_{n1}}  \nnn
 &\quad + b D \mathcal C  \frac{1}{k^2}\rbr{ j_{2n}(k\bar r_z)-\delta_{n0}}
		 + \rbr{ \mathcal B\mathcal C   -\mathcal A\mathcal D}   \frac{ 1}{k^3}\rbr{j_{2n}'(k\bar r_z)+\frac 13 k\bar r_z\delta_{n0}-\frac{2}{15}k\bar r_z\delta_{n1}} \nnn
&\quad  -  \rbr{ \mathcal B \mathcal D +    D f
		 \mathcal C }   \frac{1}{k^2}\rbr{ j_{2n}''(k\bar r_z)+\frac 13\delta_{n0}-\frac{2}{15}\delta_{n1}}- \frac{1}{k}  b D \mathcal D  j_{2n}'(k\bar r_z)+\frac{1}{k} D f   \mathcal D   j_{2n}'''(k\bar r_z) \bigg]\,, \label{PRso} 
\end{align}
for the IR-safe observer-source cross-power spectrum,
\begin{align}
& \frac{ P_{s\text{-}nl}^{\text{IR-safe}}( \bm k) }{P_m(k)}=\sum_{n=0}^\infty (-1)^n (4n+1)L_{2n}   \int_0^{\bar r_z}\mathrm d\bar r\, \bigg[ 	\mathcal A  \mathcal E   \frac{1}{k^4}\rbr{j_{2n}(\Delta x_z)-\rbr{1-\frac 16\Delta x_z^2}\delta_{n0}-\frac{1}{15}\Delta x_z^2\delta_{n1}} 
		\nnn
&\quad +\rbr{\mathcal B \mathcal E   -\mathcal A \mathcal F}   \frac{1}{k^3}\rbr{j_{2n}'(\Delta x_z)+\frac 13 \Delta x_z\delta_{n0}-\frac{2}{15}\Delta x_z\delta_{n1}} + b D  \mathcal E   \frac{1}{k^2}\rbr{j_{2n}(\Delta x_z)-\delta_{n0}}   \nnn
&\quad + \mathcal A  \mathcal G   \frac{ 1}{k^2}\rbr{j_{2n}(\Delta x_z) +  j_{2n}''(\Delta x_z) - \frac 23\delta_{n0}-\frac{2}{15}\delta_{n1}} +  \mathcal B \mathcal G   \frac{ 1}{k}\rbr{j_{2n}'(\Delta x_z) +  j_{2n}'''(\Delta x_z)} 
		\nnn
&\quad - \rbr{ D f  \mathcal E  + \mathcal B\mathcal F } \frac{1}{k^2}\rbr{j_{2n}''(\Delta x_z)+\frac 13\delta_{n0}-\frac{2}{15}\delta_{n1}}  
 -  b D \mathcal F  \frac{1}{k}   j_{2n}'(\Delta x_z) 
 \nnn
 &\quad + b D   \mathcal G \rbr{  j_{2n}(\Delta x_z) +  j_{2n}''(\Delta x_z)}	-D f  \mathcal G \rbr{ j_{2n}''(\Delta x_z) + j_{2n}''''(\Delta x_z)}  +  D f \mathcal F  \frac{1}{k}  j_{2n}'''(\Delta x_z)\bigg]\,, \label{PRsnl}
\end{align}
for the IR-safe source-non local cross power spectrum, and
\begin{align}
&\frac{P_{o\text{-}nl}^{\text{IR-safe}}( \bm k)}{P_m(k)} 
= \sum_{n=2}^\infty (-1)^n (4n+1)L_{2n}   \int_0^{\bar r_z}\mathrm d\bar r\,       \bigg[ \mathcal C\mathcal E    \frac{1}{k^4}  \rbr{j_{2n}(k\bar r)-\rbr{1-\frac 16k^2\bar r^2}\delta_{l0}-\frac{1}{15}k^2\bar r^2\delta_{l1}} \nnn
& \quad- \rbr{ \mathcal D \mathcal E - \mathcal C\mathcal F }  \frac{1}{k^3}  \rbr{j_{2n}'(k\bar r)+\frac 13k\bar r\delta_{l0}-\frac{2}{15}k\bar r\delta_{l1}}-  \mathcal D \mathcal F   \frac{1}{k^2} \rbr{j_{2n}''(k\bar r)+\frac 13\delta_{l0}-\frac{2}{15}\delta_{l1}}\nnn
 &  \quad + 
	\mathcal C	 \mathcal G     \frac{1}{k^2} \rbr{  j_{2n}(k\bar r) +  j_{2n}''(k\bar r)-\frac 23\delta_{l0}-\frac{2}{15}\delta_{l1}}-\mathcal D \mathcal G \frac{1}{k}  \rbr{ j_{2n}'(k\bar r)  + j_{2n}'''(k\bar r)}\bigg] 
\,,\label{PRonl} 
\end{align}
for the IR-safe observer-non local cross power spectrum. 
The expressions for the monopole, quadrupole and hexadecapole cross power spectra and their divergent and IR-safe parts can be read off the above expressions by taking the coefficients of $L_0$, $L_2$ and $L_4$, respectively.

\newpage

\end{document}